\documentclass[utf8]{frontiersSCNS} 
\usepackage{url,hyperref,lineno,microtype,subcaption}
\usepackage[onehalfspacing]{setspace}
\usepackage{tablefootnote}

	\usepackage{booktabs}
	\usepackage{amsmath}
	\usepackage{textcomp}
	\usepackage{multirow}
	\usepackage[euler]{textgreek}
	\usepackage{soul}

   \graphicspath{{./figures/}}

\def\keyFont{\fontsize{8}{11}\helveticabold }
\def\firstAuthorLast{Katsenou {et~al.}} 
\def\Authors{Angeliki Katsenou\,$^{*}$, Fan Zhang, Mariana Afonso, Goce Dimitrov, and David R. Bull\,}
\def\Address{Visual Information Laboratory, University of Bristol, Department of Electrical and Electronic Engineering, Bristol, UK}
\def\corrAuthor{Angeliki Katsenou}

\def\corrEmail{angeliki.katsenou@bristol.ac.uk}

\begin{document}
\onecolumn
\firstpage{1}

\title[Introducing BVI-CC Dataset]{BVI-CC: A Dataset for Research on Video Compression and Quality Assessment} 

\author[\firstAuthorLast ]{\Authors} 
\address{} 
\correspondance{} 

\extraAuth{}

\maketitle

\begin{abstract}
The video technology scenery has been very vivid over the past years, with novel video coding technologies introduced that promise improved compression performance over state-of-the-art technologies. Despite the fact that a lot of video datasets are available, representative content of the wide parameter space along with subjective evaluations of variations of encoded content from an unpartial end is required. In response to this requirement, this paper features a dataset, the BVI-CC. Three video codecs were deployed to create the variations of the encoded sequences: High Efficiency Video Coding (HEVC) Test Model (HM), AOMedia Video 1 (AV1), and Versatile Video Coding (VVC) Test Model (VTM). Nine source video sequences were carefully selected to offer both diversity and representativeness in the spatio-temporal domain. Different spatial resolution versions of the sequences were created and encoded by all three codecs at pre-defined target bit rates. The compression efficiency of the codecs was evaluated with commonly used objective quality metrics, and the subjective quality of their reconstructed content was also evaluated through psychophysical experiments. Additionally, an adaptive bit rate (convex hull rate-distortion optimization across spatial resolutions) test case was assessed using both objective and subjective evaluations. Finally, the computational complexities of the tested codecs were examined. 
All data have been made publicly available as part of the dataset, which can be used for coding performance evaluation and video quality metric development.

\tiny
 \keyFont{ \section{Keywords:} video codec dataset, codec comparison, HEVC, AV1, VVC, objective quality assessment, subjective quality assessment.} 
\end{abstract}

\section{Introduction}
\label{sec:intro}
Video technology is ubiquitous in modern life, with wired and wireless video streaming, terrestrial and satellite TV, Blu-ray players, digital cameras, video conferencing and surveillance all underpinned by efficient signal representations. It was predicted that, by 2022, 82\% (approximately 4.0ZB) of all global internet traffic per year will be video content \cite{r:cisco2}. This projected figure was probably hit earlier due to the increased use of video technologies during the Covid-19 pandemic. It is therefore a very challenging time for compression, which must efficiently encode these increased quantities of video at higher spatial and temporal resolutions, dynamic resolutions and qualities.

The last three decades have witnessed significant advances in video compression technology, from the first international video coding standard H.120 (\cite{r:h120}), to the widely adopted MPEG-2/H.262 (\cite{r:h262}), and H.264 Advanced Video Coding (H.264/AVC) (\cite{r:h264}) standards. Recently, ISO/IEC Moving Picture Experts Group (MPEG) and ITU-T Video Coding Experts Group (VCEG) have released a new video coding standard, Versatile Video Coding (VVC) (\cite{s:VVC1}), with the aim of reducing bit rates by 30\%-50\% compared to the current High Efficiency Video Coding (HEVC) standard (\cite{r:HEVC}). In parallel, the Alliance for Open Media (AOMedia) have developed royalty-free open-source video codecs to compete with MPEG standards. The first AOMedia Video 1 (AV1) codec (\cite{w:AV1,AV1specs}) has been reported to outperform its predecessor VP9, developed by \cite{w:VP9}.  
In order to benchmark these coding algorithms, their rate quality performance can be evaluated using objective and/or subjective assessment methods. Existing works, by \cite{c:Akyazi,c:Grois,c:SeixasDias,c:Guo,c:Zabrovskiy,c:Katsavounidis, nguyen_marpe_2021}, have reported comparisons for contemporary codecs, with perplexing results and conclusions, mainly due to the use of different coding configurations. Also, most of these studies are solely based on objective quality assessment. Finally, the majority of these works do not publicly release the produced data.

The significant impact of data availability has always been important in video technology research and has become even more crucial over the past years due to the deployment of machine learning and deep learning methods (see~\cite{BVI-DVC}). Furthermore, existing datasets lack of variety in the encoded versions of the raw sequences, as the majority offers only H.264 (e.g. VQEGHD3 (\cite{r:vqegHD}), LIVE (\cite{LIVE}), NFLX-P and VMAF+ (\cite{w:VMAF})) or HEVC (e.g. BVI-HD (\cite{j:Zhang7}), BVI-Texture (\cite{w:bviTex}), BVI-SynTex (\cite{BVI-SynTex})) encodings.

In this context, this paper presents a video dataset, referred to as BVI-CC, which comprises a complete set of 306 encodings using VVC, HEVC, and AV1, on nine representative source sequences typically used by the standardisation bodies. To this end, a controlled set of three experiments were designed taking into consideration the codecs corresponding common test conditions. 
The source sequences native spatial resolution is Ultra High Definition (UHD), 3840$\times$2160. Additionally to the UHD resolution, the sequences were spatially downscaled to 1920$\times$1080, 1280$\times$720, and 960$\times$544 resolution and three experiments were designed. Two experiments were traditionally configured using constant resolution, one at UHD and one at HD. The third experiment was designed on an adaptive bit rate use case implementing the Dynamic Optimizer (DO) approach\footnote{Here only convex hull rate-distortion optimisation within each shot is employed.} as described by \cite{w:DO} and implemented by \cite{c:Zhang24} (up to FHD resolution only).  
This work provides a comprehensive extension of our previous work (\cite{c:Zhang24}), where only AV1 and HEVC results were presented based on the DO approach. BVI-CC is complemented by data collected during the quality assessment: (i) anonymised opinion scores from psychovisual experiments in labs and (ii) values of six commonly used objective quality metrics. BVI-CC dataset is publicly available after request (see~\cite{BVI-CC}) and can be utilised either for research on video compression and research on image/video quality assessment.

The rest of this paper is organised as follows. Section \ref{sec:background} briefly reviews the history of video coding and related work on codec comparison. Section \ref{sec:database} presents the selected source sequences and the coding configurations employed in generating various compressed content. In Section \ref{sec:exp}, the conducted subjective experiments are described in detail, while the evaluation results through both objective and subjective assessment are reported and discussed in Section \ref{sec:results}. Finally, Section \ref{sec:conclusions} outlines the conclusion and future work. 

\section{Background}
\label{sec:background} 

This section provides a brief overview of video coding standards and reports on existing datasets for research on video compression and quality assessment.

\subsection{Video Coding Standards and Technologies}
Video coding standards normally define the syntax of bitstream and the decoding process, while encoders generate standard-compliant bitstream and thus determine compression performance. Each generation of video coding standard comes with a reference test model, such as HM (HEVC Test Model) for HEVC, which can be used to provide a performance benchmark. 
H.264/MPEG-4-AVC (\cite{r:h264}) was launched in 2004, and is still the most prolific video coding standard, despite the fact that its successor, H.265/HEVC (\cite{r:HEVC}) finalised in 2013, provides enhanced coding performance. In 2020, the first version of the latest video coding standard, Versatile Video Coding (VVC) (\cite{s:VVC1}), has been finalised, which can achieve 30\%-50\% coding gain over H.265/HEVC, supporting immersive formats (360\textdegree~videos) and higher spatial resolutions, up to 16K.  

Alongside recent MPEG standardisation, there has been increasing activity in the development of open-source royalty-free video codecs, particularly by the Alliance for Open Media (AOMedia), a consortium of video-related companies. VP9 (\cite{w:VP9}), which was earlier developed by Google to compete with MPEG, provided a basis for AV1 (AOMedia Video 1) (\cite{w:AV1, AV1specs}), which was released in 2018. \cite{AV1specs} set as AV1's main goal to provide an open source and royalty-free video coding format that substantially outperforms its primary market competitors in compression efficiency and concurrently achieves a practical decoding complexity, is optimized for hardware feasibility and scalability on modern devices. Since its release, its pool of contributors has expanded and they have provided significant updates towards its goal. The AOM contributors are currently working towards developing the next generation, AV2 (\cite{AV2blog}). For further details on existing video coding standards and formats, the readers are referred to \cite{b:Bull,b:Wien,b:Ohm}.

The performance of video coding algorithms is usually assessed by comparing their rate-distortion (RD) or rate-quality (RQ) performance on various test sequences. According to \cite{r:BT500}, the selection of test content is important and  should provide a diverse and representative coverage of the video parameter space. Objective quality metrics or/and subjective opinion measurements are normally employed to assess compressed video quality. The overall RD or RQ performance difference between codecs can be then calculated using Bj{\o}ntegaard delta metrics for objective quality metrics according to \cite{r:Bjontegaard} or SCENIC developed by \cite{j:Hanhart} for subjective assessments. 

Most recent literature on video codec comparisons has focused on performance evaluations between MPEG codecs (H.264/AVC, HEVC, and VVC) and royalty-free (VP9 and AV1) codecs (\cite{c:Akyazi,c:Grois,j:MMSPSVD,c:SeixasDias, nguyen_marpe_2021}) and on their application in adaptive video steaming services \cite{c:Guo,c:Zabrovskiy,c:Katsavounidis}. However, the results presented are acknowledged to be highly inconsistent(\cite{nguyen_marpe_2021}), mainly due to the different configurations employed across codecs. 

Our work contributes towards a fair codec comparison by releasing publicly a dataset that includes objective and subjective codec comparisons in three different use cases: encoding at FHD and UHD resolution and encoding within the framework of adaptive streaming.

\subsection{Datasets for Video Compression Purposes}

The literature is rich of video datasets developed for many different purposes, mainly for computer vision related tasks such as object detection, action recognition, summarization, etc. These datasets, however, are not suitable for research in video compression. The main reason is that those have been designed for training deep network or models to infer very specific information. For example, the EPIC-Kitchens dataset (\cite{Damen2021PAMI}), contains scenes of daily activities in the kitchen, e.g. slicing bread, peeling carrots, stirring soup, etc. Therefore, the content bears significant similarities by repeating specific patterns in similar although diverse set-ups. Not providing a wide range of scenes with spatial and temporal information, this type of datasets cannot adequately represent video content and, thus, form a basis for a fair comparison of video compression algorithms and/or video quality metrics. Furthermore, the datasets published for computer vision tasks include already encoded versions (usually H.264-based encodings) as exported automatically from the video recording device. Although this type of content resembles the features of user generated content (UGC), a large portion of the videos streamed are coming from the creative industry. Thus research on video technology additionally requires pristine videos as exported from post-production.

In Table~\ref{tab:SotADatasets}, a selection of commonly use datasets for video compression research is listed along with some of their basic features.
Some of these datasets consist of pristine sequences (those that include raw (uncompressed) video sequences in the dataset) and others of UGC content (no raw sequences available as in KonViD-1K (\cite{konvid1k}), YouTube-UGC (\cite{YouTube-UGC}), DVL2021 (\cite{XING2022103374})). Most of the existing datasets with pristine content, only include encoded sequences with one codec, usually H.264 (e.g. VQEGHD3 (\cite{r:vqegHD}), LIVE (\cite{LIVE}), NFLX-P and VMAF+ (\cite{w:VMAF})) or HEVC (e.g. BVI-HD (\cite{j:Zhang7}), BVI-Texture (\cite{w:bviTex}), BVI-SynTex (\cite{BVI-SynTex})). 

From Table~\ref{tab:SotADatasets}, it is evident that there is no dataset available that offers variations of encoded sequences based on different state-of-the-art codecs. Furthermore, to the best of our knowledge, there is no other dataset available that offers encoded sequences based on VVC and AV1. This is a very important contribution of the introduced dataset, BVI-CC, as it could facilitate research on video compression and comparison across these three different codecs. It is also noticeable that although almost all datasets provide opinion scores (OS) data, most of the datasets do not provide computed values of objective quality metrics (QMs).

\begin{table*}[htb]
	\centering
	\footnotesize
	\caption{Selection of State-of-the-Art Datasets for Video Research Purposes.}
	\label{tab:SotADatasets}
	
	\begin{tabular}{l | cccccc}
		\toprule
		Dataset  &Resolution  & Raw & Encoded & Codec & QMs & OS \\
		\midrule 
		VQEGHD3 (\cite{r:vqegHD})& 1080p& 13&72&MPEG2/H264& - & \checkmark \\
		LIVE (\cite{LIVE})& 1080p& 10& 150& H.264& - & \checkmark\\
		MCL-V (\cite{MCL-V}) & 1080p&12 &96 &H.264 & -& \checkmark\\
		BVI-HD (\cite{j:Zhang7}) & 1080p & 32 & 192 & HEVC &  -  & \checkmark \\ 
		BVI-Texture (\cite{w:bviTex}) & 1080p &  20 & 80 & HEVC &  -  & \checkmark \\ 
		BVI-SynTex (\cite{BVI-SynTex}) & 1080p &  49 & 196 & HEVC &  \checkmark & \checkmark \\ 
		BVI-DVC (\cite{BVI-DVC})& up to 2160p&  800 & - & - &  - & - \\ 
		100-4K (\cite{KatsenouPCS2019})& 2160p&  100 & - & - &  - & - \\ 
		VMAF+ (\cite{w:VMAF}) &  up to 1080p& 23& 230 & H.264& -& \checkmark\\
		NFLX-P (\cite{w:VMAF})& up to 1080p& 9& 70& H.264 & -& \checkmark\\
		YouTube-UGC(\cite{YouTube-UGC})  & up to 2160p & - & $\sim$ 1,500& H.264, VP9\footnotemark & \checkmark & \checkmark\\
		KonViD-1K (\cite{konvid1k})& 540p& -& 1,200 & H.264& -& \checkmark\\
		DVL2021 (\cite{XING2022103374})  &2160p &  - & 206 & - & - &\checkmark\\
		\bottomrule
	    \textbf{BVI-CC} & up to 2160p & 9 & 306 & HEVC, VVC, AV1 & \checkmark &\checkmark\\
		\bottomrule
	\end{tabular}

\end{table*}
\footnotetext{Only for a subset, for the categories Gaming, Sports, and Vlog videos.} 

\section{Test Content and Codec Configurations}
\label{sec:database} 
This section describes the selection of source sequences and the different codec configurations used to generate their various compressed versions.  

\subsection{Source Sequence Selection}
\begin{figure}[htbp]
\scriptsize
\begin{minipage}[b]{0.325\linewidth}
\centering
\centerline{\includegraphics[width=\linewidth]{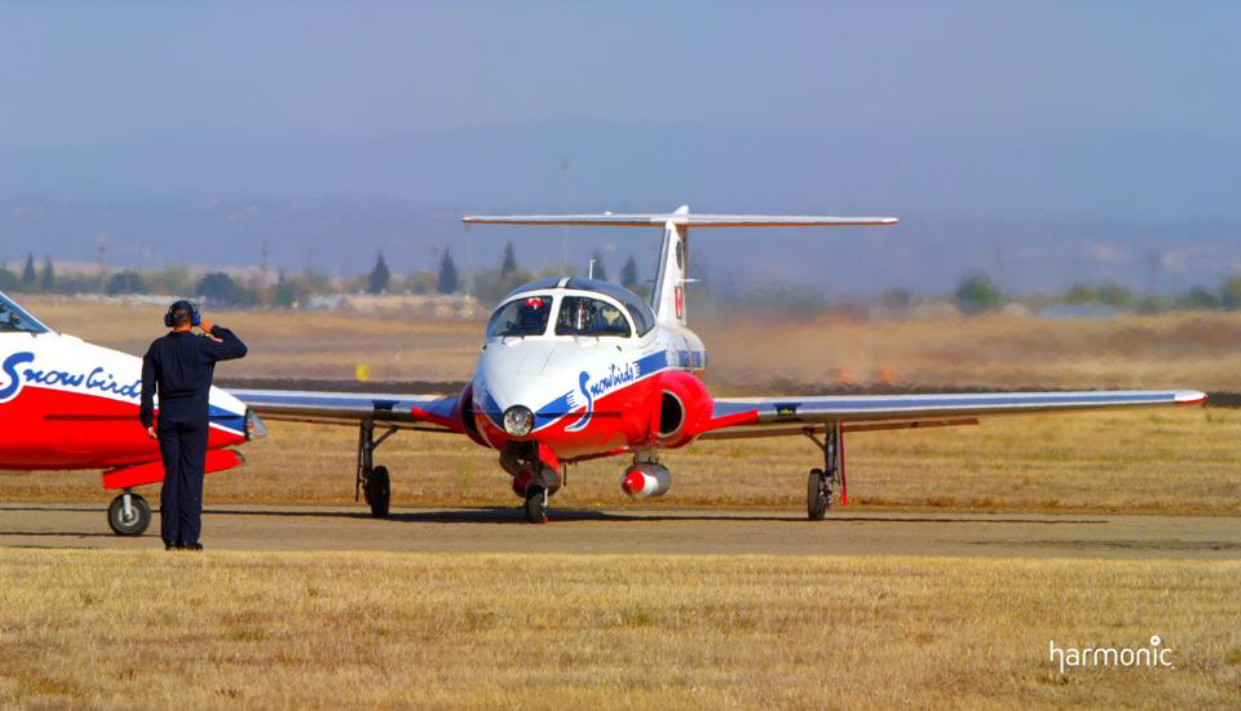}}
\centerline{V1: AirAcrobatic}
\vspace{.5em}
\end{minipage}
\begin{minipage}[b]{0.325\linewidth}
\centering
\centerline{\includegraphics[width=\linewidth]{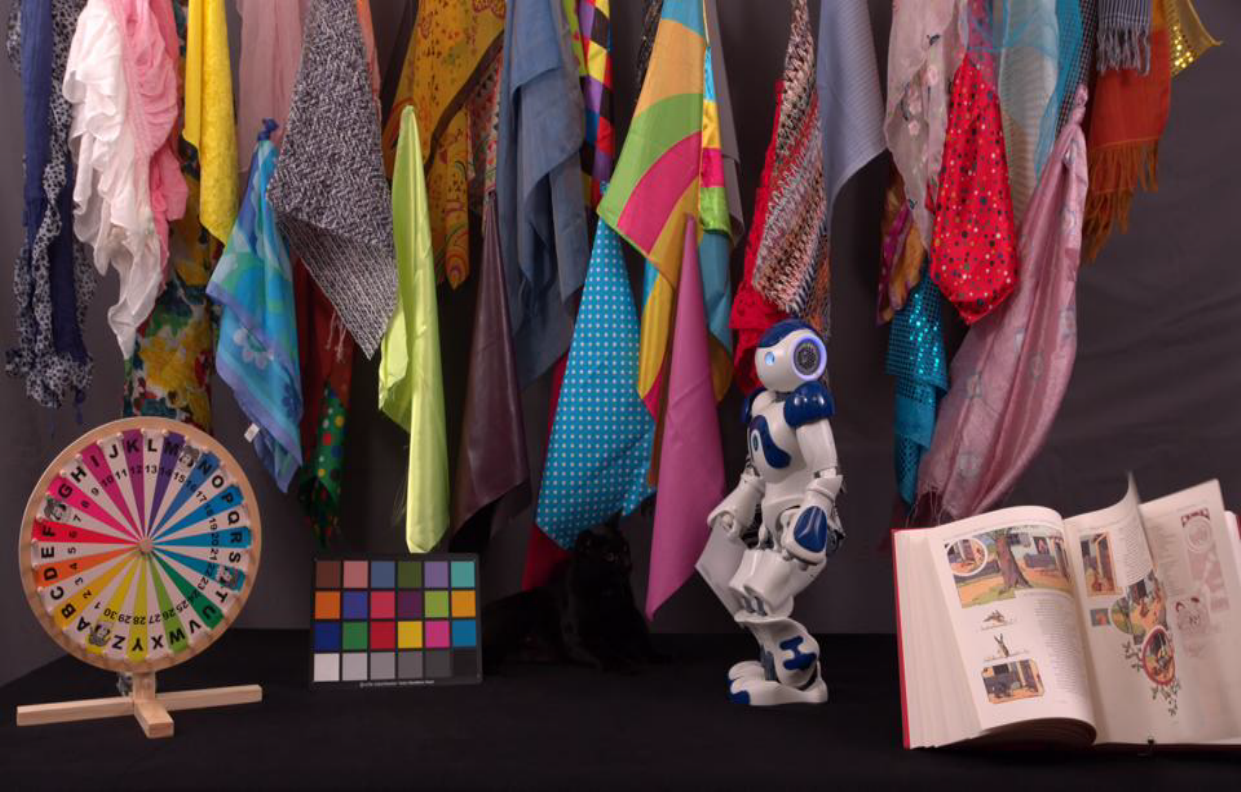}}
\centerline{V2: CatRobot}
\vspace{.5em}
\end{minipage}
\begin{minipage}[b]{0.325\linewidth}
\centering
\centerline{\includegraphics[width=\linewidth]{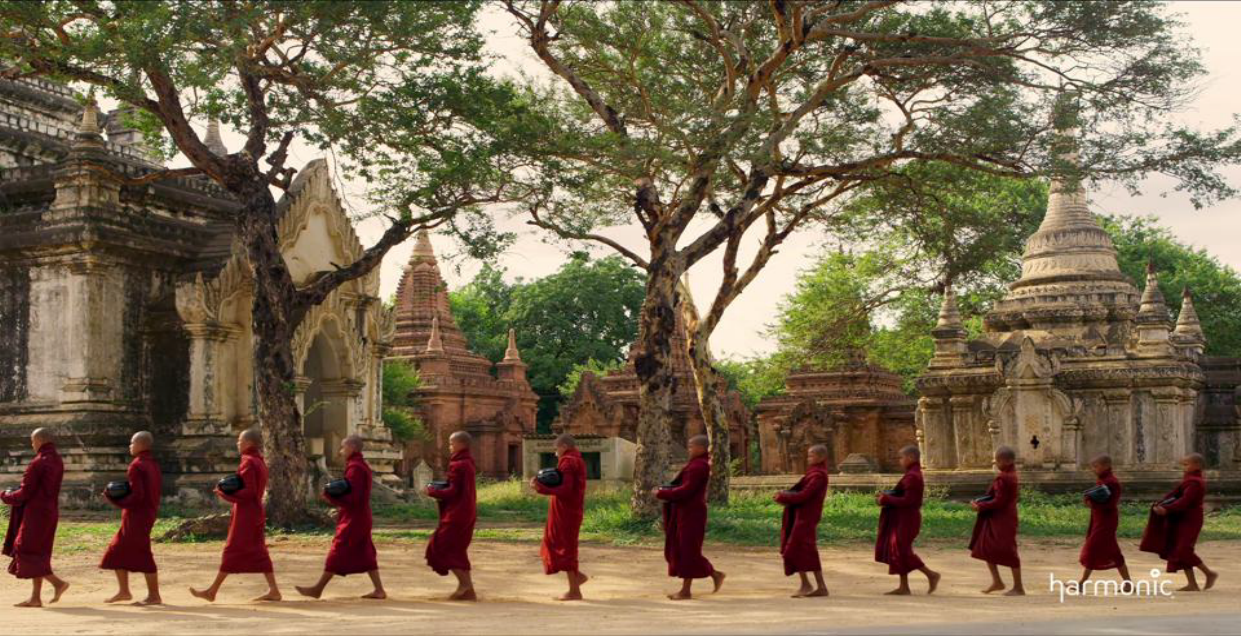}}
\centerline{V3: Myanmar}
\vspace{.5em}
\end{minipage}

\begin{minipage}[b]{0.325\linewidth}
\centering
\centerline{\includegraphics[width=\linewidth]{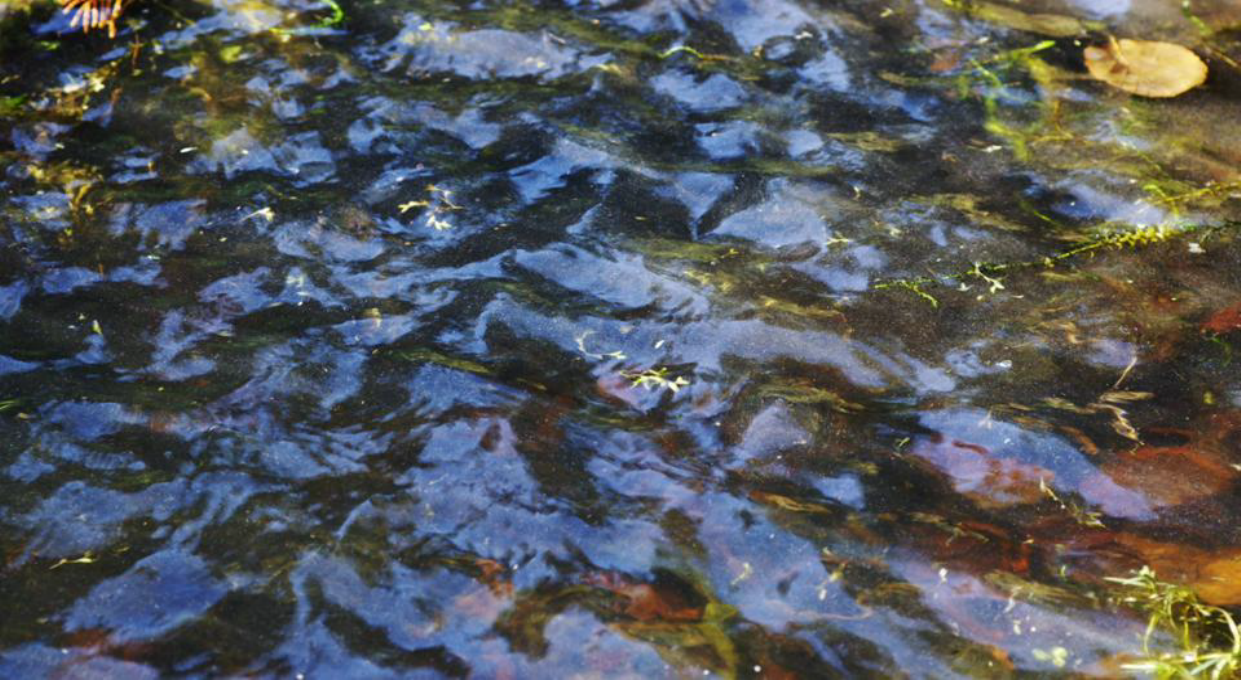}}
\centerline{V4: CalmingWater}
\vspace{.5em}
\end{minipage}
\begin{minipage}[b]{0.325\linewidth}
\centering
\centerline{\includegraphics[width=\linewidth]{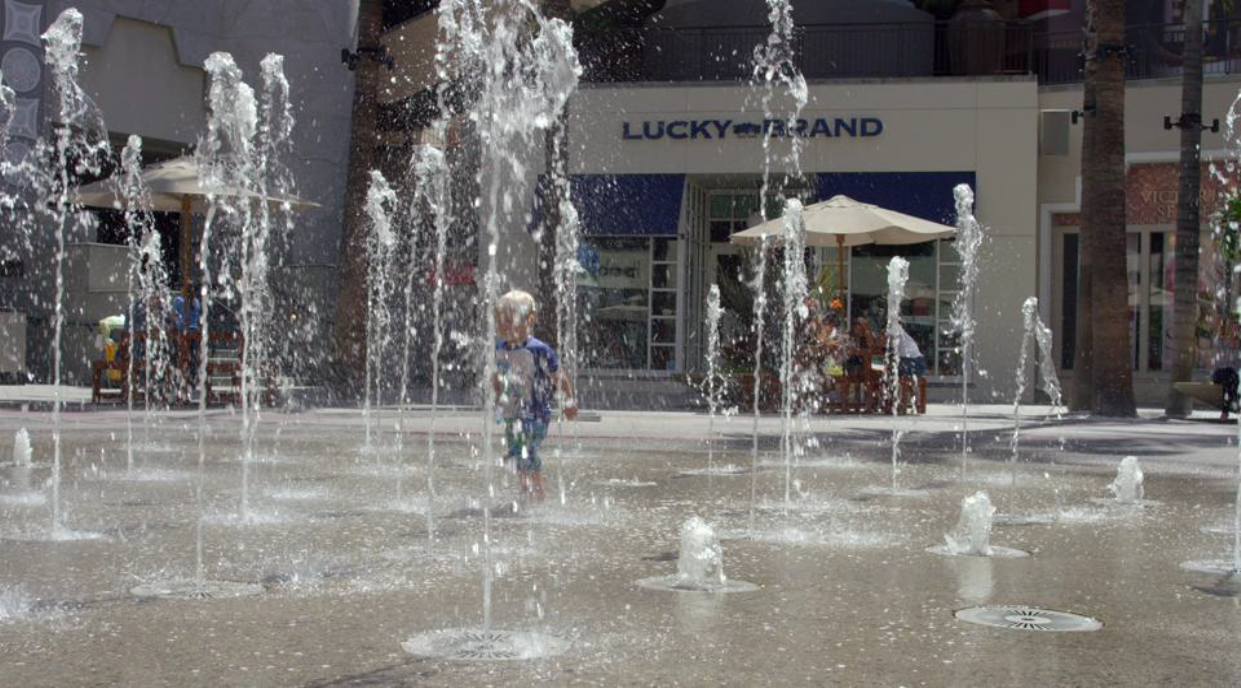}}
\centerline{V5: ToddlerFontain}
\vspace{.5em}
\end{minipage}
\begin{minipage}[b]{0.325\linewidth}
\centering
\centerline{\includegraphics[width=\linewidth]{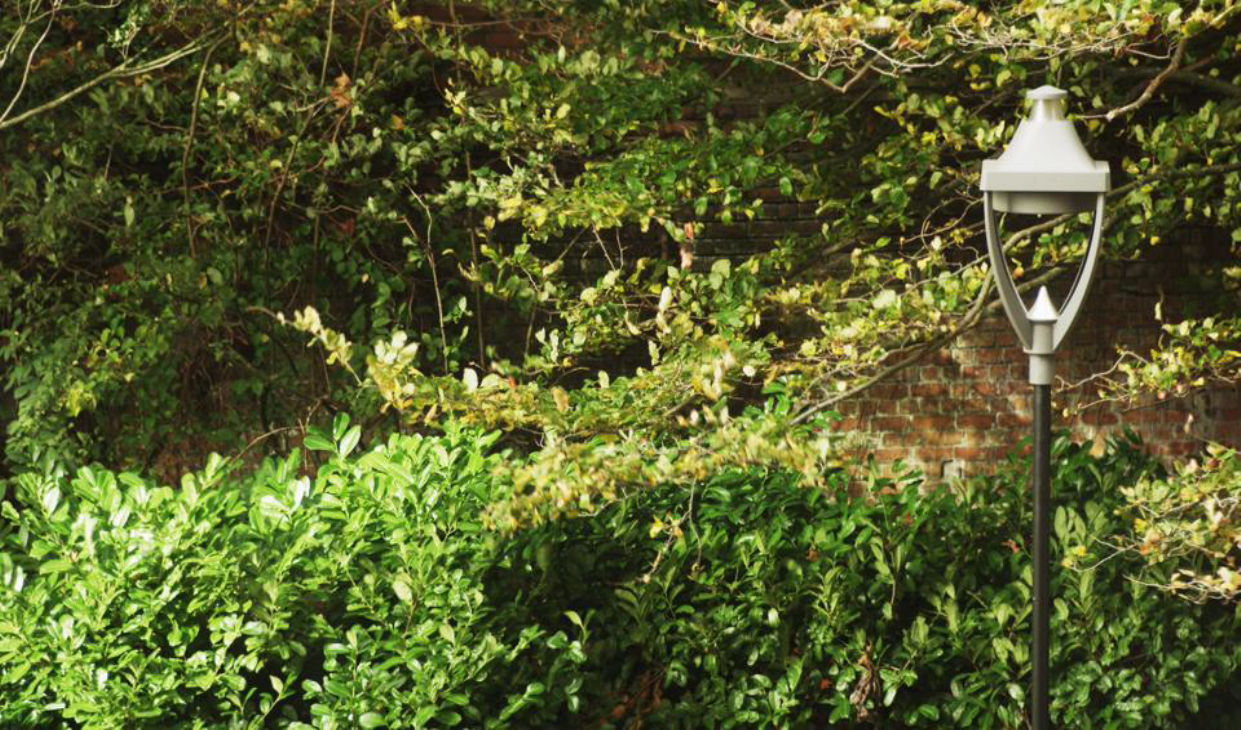}}
\centerline{V6: Lampleaves}
\vspace{.5em}
\end{minipage}

\begin{minipage}[b]{0.325\linewidth}
\centering
\centerline{\includegraphics[width=\linewidth]{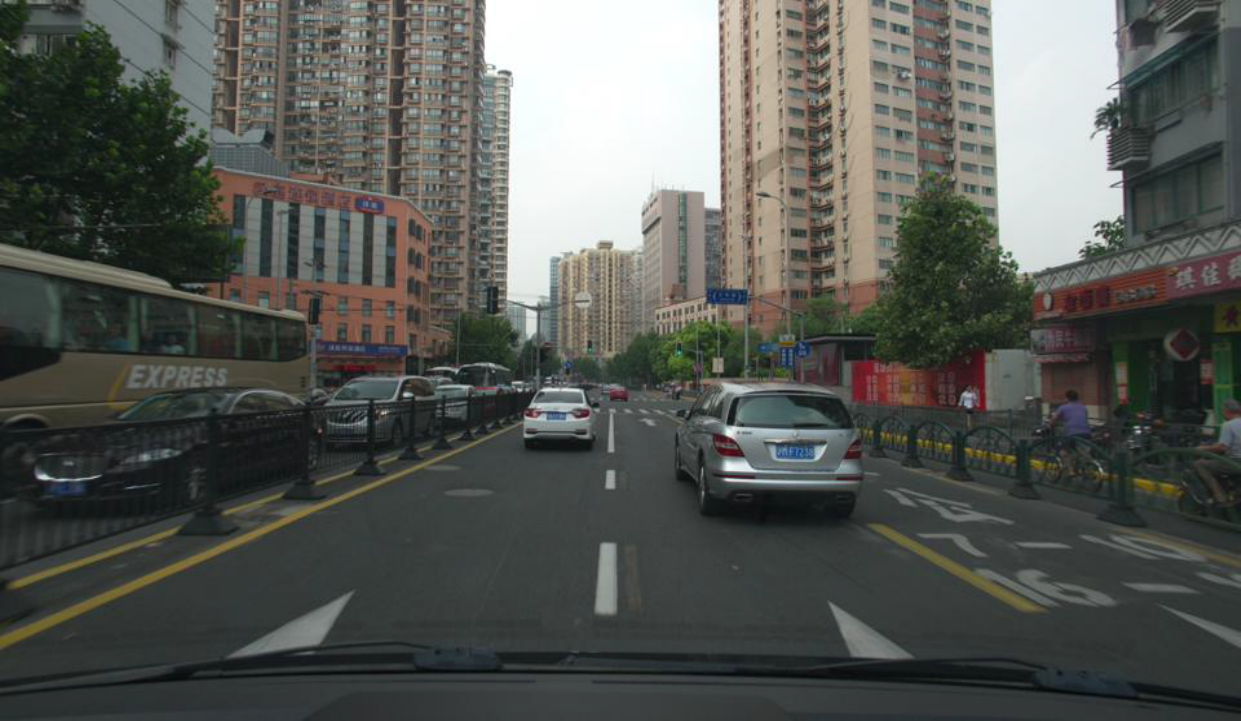}}
\centerline{V7: DaylightRoad}
\end{minipage}
\begin{minipage}[b]{0.325\linewidth}
\centering
\centerline{\includegraphics[width=\linewidth]{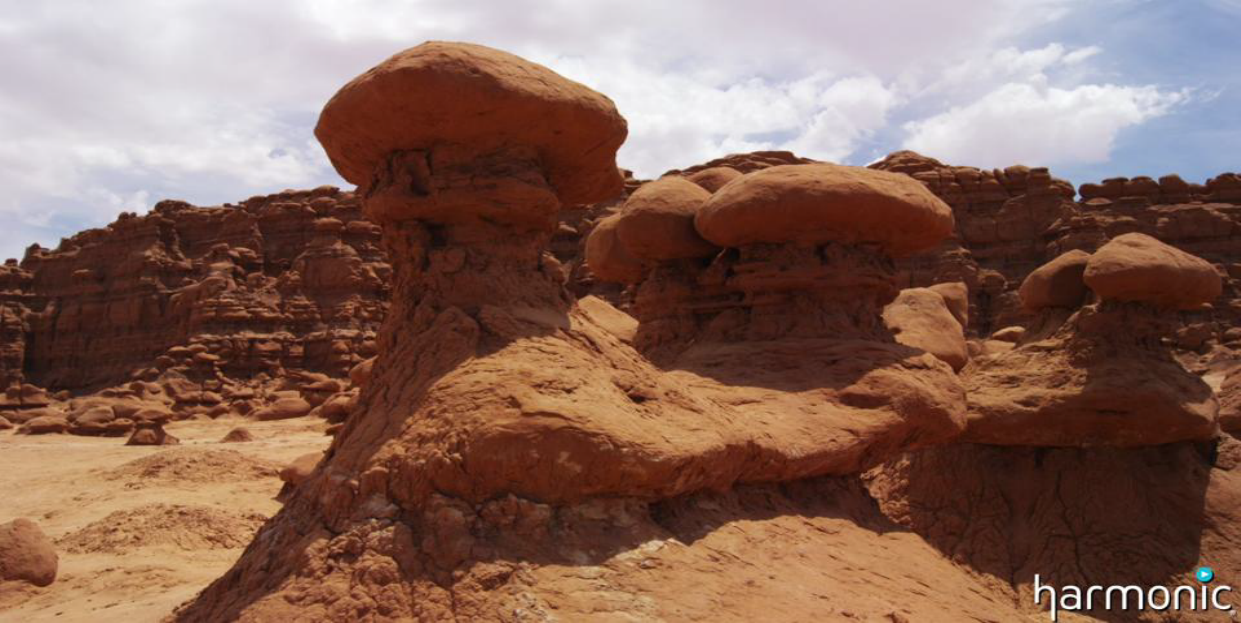}}
\centerline{V8: RedRock}
\end{minipage}
\begin{minipage}[b]{0.325\linewidth}
\centering
\centerline{\includegraphics[width=\linewidth]{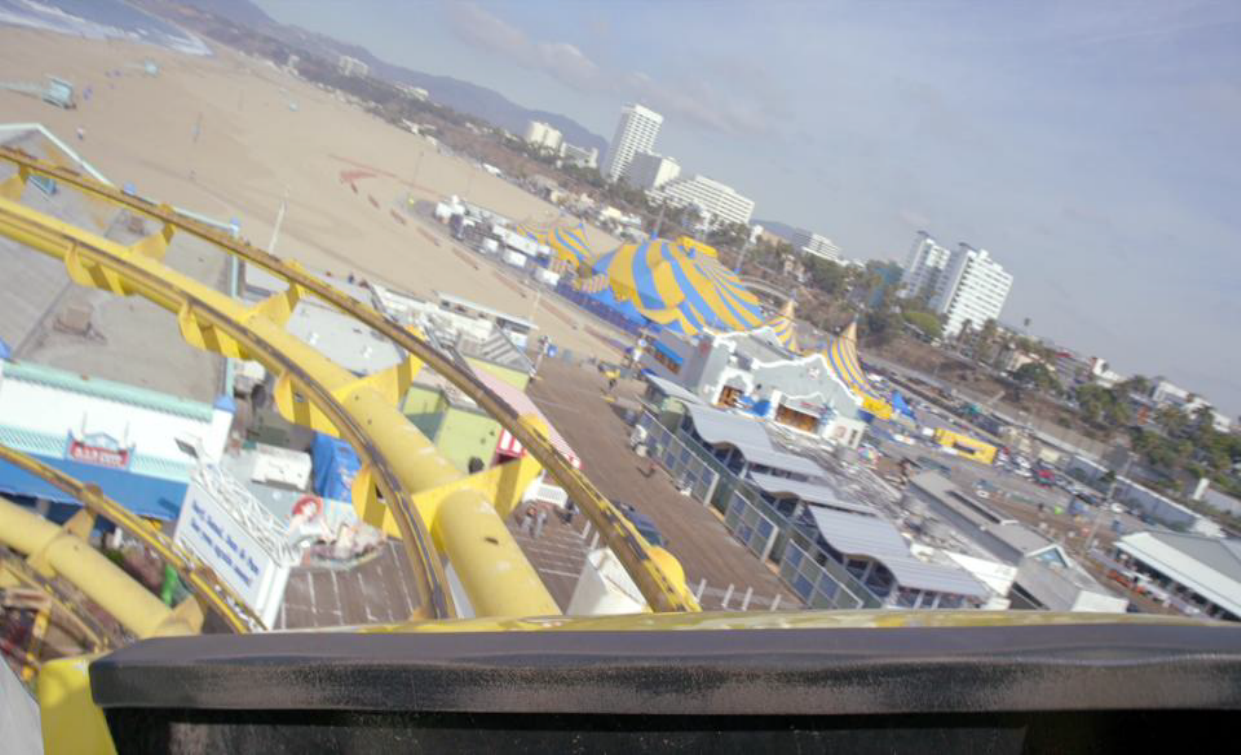}}
\centerline{V9: RollerCoaster}
\end{minipage}
\caption{Sample video frames from the selected source sequences.}
\label{fig:samples}
\end{figure}

\begin{figure}[htbp]
\centering
\includegraphics[width=.5\linewidth]{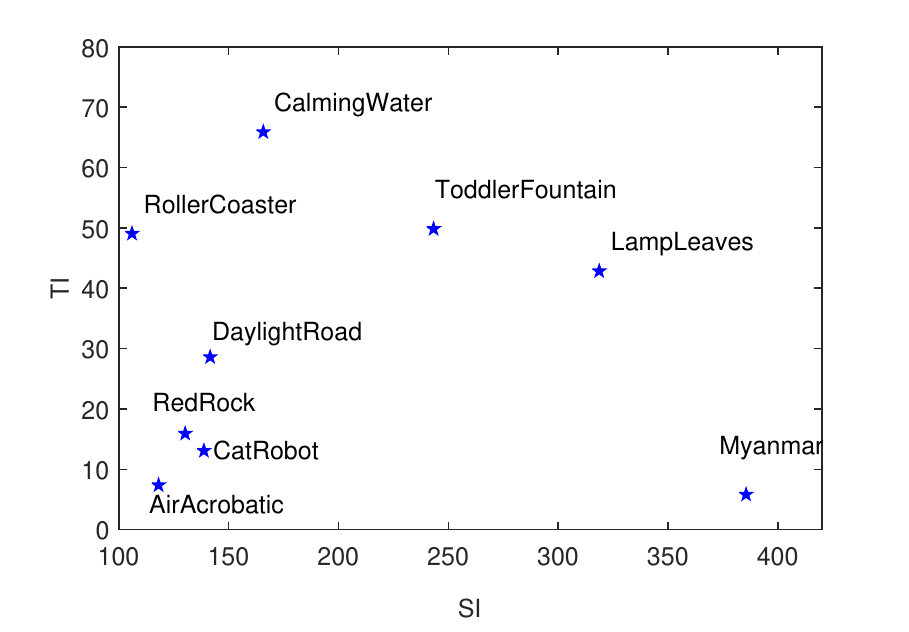}
\caption{Scatter plot of SI and TI for the selected source sequences.}
\label{fig: SITI}
\end{figure}

Nine source sequences were selected from \cite{w:Harmonic}, BVI-Texture (\cite{w:bviTex}) and JVET (Joint Video Exploration Team) CTC (Common Test Conditions) datasets. Each sequence is progressively scanned, at Ultra High Definition (UHD, 3840$\times$2160) resolution, with a frame rate of 60 frames per second (fps) and without scene-cuts. All were truncated from their original lengths to five seconds (rather than the recommended 10 seconds in ITU standard \cite{r:BT500}). This reflects the recommendations of a recent study on optimal video duration for subjective quality assessment by \cite{j:Zhang4,j:Zhang6}. Sample frames from the selected nine source sequences alongside clip names and indices are shown in Fig. \ref{fig:samples}. The dataset includes three sequences with only local motion (without any camera motion, V1-V3), three sequences with dynamic textures (for definitions see \cite{j:Zhang}, V4-V6), and three with complex camera movements (V7-V9). The coverage of the video parameter space is confirmed in Fig.~\ref{fig: SITI}, where the Spatial and Temporal Information of the dataset (SI and TI) as defined by \cite{j:Winkler1} are plotted.  

In order to investigate coding performance for different resolutions and within an adaptive streaming framework, three spatial resolution groups were generated from the source sequences: (A) UHD (3840$\times$2160) only, (B) HD (1920$\times$1080) only, and (C) HD-Dynamic Optimizer (HD-DO). For group C, coding results for three different resolutions (1920$\times1080$, 1280$\times$720, and 960$\times$544) and with various quantisation parameters (QPs) were firstly generated. The reconstructed videos were then up-sampled to HD resolution (in order to provide a basis for comparison with the original HD sequences). Here, spatial resolution re-sampling was implemented using Lanczos-3 filters designed by \cite{j:Duchon}. The rate points with optimal rate-quality performance based on Video Multimethod Assessment Fusion (VMAF) (\cite{w:VMAF}) were selected across the three tested resolutions for each target bit rate and codec. This process is repeated to create the entire convex hull in the DO approach (\cite{w:DO}). The resulting selected resolutions across the set of target bit rates are reported in Table~\ref{tab:DORes}.

\begin{table*}[htb]
	\centering
	\footnotesize
	\caption{Resolution selections per sequence after applying the DO methodology on AV1 and HM for resolution group C.}
	\label{tab:DORes}
	
	\begin{tabular}{l | ccccc | ccccc }
		\toprule
		 \multirow{2}{*}{Sequences}   &\multicolumn{5}{c|}{HM} & \multicolumn{5}{c}{AV1} \\
		\cmidrule{2-11}
		 	 &   R1 & R2 & R3 & R4 & R5  &		R1 & R2 & R3 & R4 & R5\\
		\midrule 
		V1: AirAcrobatic & 720p & 1080p& 1080p& 1080p& 1080p& 		720p& 720p& 720p& 720p& 720p\\
		\midrule 
		V2: CatRobot&	 1080p & 1080p& 1080p& 1080p& 1080p&		720p& 720p& 1080p& 1080p& 1080p\\
		\midrule 
		V3: Myanmar  & 1080p & 1080p& 1080p& 1080p& 1080p& 	720p& 1080p& 1080p& 1080p& 1080p\\
		\midrule 
		V4: CalmingWater	  & 544p & 544p& 544p& 544p& 720p& 	 		544p& 544p& 544p& 544p& 720p\\
		\midrule 
		V5: ToddlerFountain  	& 544p & 544p& 544p& 544p& 720p& 			544p& 544p& 544p& 544p& 720p\\
		\midrule 
		V6: LampLeaves & 720p & 720p& 1080p& 1080p& 1080p& 		544p& 544p& 720p& 720p& 720p\\
		\midrule 
		V7: DaylightRoad & 720p & 1080p& 1080p& 1080p& 1080p& 		720p& 720p& 1080p& 1080p& 1080p\\
		\midrule 
		V8: RedRock  & 720p & 720p& 1080p& 1080p& 1080p& 		544p& 720p& 720p& 720p& 720p\\
		\midrule 
		V9: RollerCoaster & 544p & 544p& 720p& 720p& 1080p& 		544p& 544p& 720p& 720p& 720p\\
		\bottomrule			
	\end{tabular}
	\vspace{-.2cm}
\end{table*}

\subsection{Coding Configurations}
\label{sec:config}
The reference test models of HEVC and VVC, and their major competitor, AV1 have been evaluated in this study. Each codec was configured using the coding parameters defined in their common test conditions (see \cite{r:JCTVCAF1100,s:JVETCTC,r:AV1CTC}), with fixed quantisation parameters (rate control disabled), the same structural delay (e.g. defined as GOP size in the HEVC HM software) of 16 frames and the same random access intervals (e.g. defined as IntraPeriod in the HEVC HM software) of 64 frames. The actual codec versions and configuration parameters are provided in Table \ref{tab:codecs}.

Different target bit rates were pre-determined for each test sequence and for each resolution group (four points for resolution group A and B, and five for HD-DO group), and their values are shown in Table \ref{tab:bitrate}. These were determined based on the preliminary encoding results of the test sequences for each resolution group using AV1. This decision was made because the version of AV1 employed restricted production of bitstreams at pre-defined bit rates, as only integer quantisation parameters could be used. On the other hand, for HEVC HM and VVC VTM this was easier to achieve by enabling the ``QPIncrementFrame" parameter. In order to achieve these target bit rates, the quantisation parameter values were iteratively adjusted to ensure the output bit rates were sufficiently close to the targets (within a range of $\pm 3\%$).
\begin{table*}[ht]
 \caption{The software versions and configurations of the evaluated video codecs.}
\centering 
\small
		\begin{tabular}{c | c | p{11cm}}
		\toprule
	Codec	& Version & Configuration parameters \\
		\midrule
				\midrule
HEVC HM & 16.18 & Random access configuration for Main10 profile \cite{r:JCTVCAF1100}. IntraPeriod=64 and GOPSize=16.\\
		\midrule
AOM AV1 & 0.1.0-9647-ga6fa0877f & Common settings with high latency CQP configuration \cite{r:AV1CTC}. \\
& &Other coding parameters: passes=2, cpu-used=1, kf-max-dist=64, kf-min-dist=64, arnr-maxframes=7, arnr-strength=5, lag-in-frames=16, aq-mode=0, bias-pct=100, minsection-pct=1, maxsection-pct=10000, auto-alt-ref=1, min-q=0, max-q=63, max-gf-interval=16, min-gf-interval=4 and color-primaries=bt709.\\
		\midrule
VVC VTM & 4.01 & Random access configuration \cite{s:JVETCTC}. \\
& & IntraPeriod=64 and GOPSize=16.  \\
\bottomrule
\end{tabular} 
\label{tab:codecs}
\end{table*}

\begin{table*}[ht]
 \caption{Pre-determined target bit rates for all test sequences in three resolution groups.}
\centering
\footnotesize
		\begin{tabular}{l | r r r r |r r r r| r r r r r }
		\toprule
	\multirow{4}{*}{Sequence} & \multicolumn{13}{c}{Target bit rates (kbps)} \\ 
	\cmidrule{2-14}	
	& \multicolumn{4}{c|}{Resolution Group A (UHD only)} & \multicolumn{4}{c|}{Resolution Group B (HD only)} & \multicolumn{5}{c}{Resolution Group C (HD-DO)} \\
		\cmidrule{2-14}	
	& R1 & R2 & R3 & R4 & R1 & R2 & R3 & R4 & R1 & R2 & R3 & R4 & R5 \\ 	
	\midrule
		\midrule 
	V1: AirAcrobatic 		& 1300 	&	2250	&4700		&9270			&550		& 920		& 1850   &3400		& 305    &     575    &     940   &     1770    &    3350 \\
		\midrule 
	V2: CatRobot				& 3170  & 5450  &8450   &14500    &1480   & 2200  & 3250   &5440		& 910    &    1500    &    2200   &     3250    &    5500\\
		\midrule 
	V3: Myanmar 				& 11000 & 21100 &33500 	&46000		&3450   & 5500  & 8200   &10800		& 2100   &     3450   &     5500  &      8150   &    10800\\
		\midrule 
	V4: CalmingWater 		& 10100 & 19250 &30000  &50000 		&3100   & 6400  &12000   &21000		& 1140   &     3050   &     6550  &     12200   &    20500\\
		\midrule 
	V5: ToddlerFountain & 13180	& 27000	&38350  &69800 		&6150   & 13500 &21500   &34500		& 2900   &     5900   &    13200  &     20500   &    34900\\
		\midrule 
	V6: LampLeaves 			& 14550 & 26460 &43900  &69800 		&8100   & 14200 &20500   &33000		&  5030  &      8100  &     14000 &      20500  &     33500\\
		\midrule 
	V7: DaylightRoad 		& 2650  & 4450  & 7050  &12170 		&1220   &  1800 & 3000   & 5300		& 810    &    1220    &    1800   &     3000    &    5300\\
		\midrule 
	V8: RedRock 				& 1500  & 2600  & 4000  &6380 		&	680   &  1000 & 1650   & 2500   & 460    &     650    &    1020   &     1600    &    2450\\
		\midrule 
	V9: RollerCoaster 	& 1750  & 2880  & 4600  &7350 		&880    &  1480 & 2270   & 3564   &  550   &      850   &     1480  &      2280   &     3580\\
\bottomrule
\end{tabular} 
\label{tab:bitrate}
\end{table*} 

\subsection{Summary}
In summary, a total number of 306 distorted sequences were produced: there are 108 (9 source sequences $\times$ 4 rate points $\times$ 3 codecs) for Resolution Group A (UHD only), 108 (9$\times$4$\times$3) for Resolution Group B (HD only), and 90 (9$\times$5$\times$2) for Resolution Group C (HD-DO)\footnote{We have not compared VVC with other codecs using the DO approach. This is mainly due to the high computation complexity of VVC and the limited computational resources that we have. Preliminary results for Group A and B have already shown the significant improvement of VVC over the other two.}.

\section{Subjective Experiments}
\label{sec:exp}
Three subjective experiment sessions were conducted separately on the test sequences in the three resolution groups. The experimental setup, procedure, test methodology and data processing approach are reported in this section. 

\subsection{Environmental Setup}
All three experiment sessions were conducted in a darkened, living room-style environment. The background luminance level was set to 15\% of the peak luminance of the monitor used (62.5 lux) \cite{r:BT500}. All test sequences were shown at their native spatial resolution and frame rates, on a consumer display, a SONY KD65Z9D LCD TV, which measures 1429$\times$804mm, with a peak luminance of 410 lux. The viewing distance was set to 121cm (1.5 times the screen height) for Resolution Group A (UHD) and 241cm (three times the screen height) for Resolution Group B (HD) and C (HD-DO), following the \cite{r:BT500} and \cite{r:P910}. The presentation of video sequences was controlled by a Windows PC running an open source software, BVI-SVQA, developed by~\cite{SVQAlink} at the University of Bristol for psychophysical experiments.

\subsection{Experimental Procedure}
\label{sec:test}
In all three experiments, the Double Stimulus Continuous Quality Scale (DSCQS) (\cite{r:BT500}) methodology was used. In each trial, participants were shown a pair of sequences twice, including original and  encoded versions. The presentation order was randomised in each trial and was unknown to each participant. Participants had unlimited time to respond to the following question (presented on the video monitor):  ``\textit{Please rate the quality (0-100) of the first/second video. Excellent--90, Good--70, Fair--50, Poor--30 and Bad--10}''. Participants  then used a mouse to scroll through the vertical scale and score (0-100) for these two videos.  The total duration of each experimental session was approximately 50 (Resolution Group A and B) or 60 (Resolution Group C) minutes, and each was split into two sub-sessions with a 10 minute break in between. Before the formal test, a training session was conducted, under the supervision of the instructor, consisting of three trials (different from those used in the formal test) to allow the participants time to familiarize.

\subsection{Participants and Data Processing}
 
A total of 62 subjects, with an average age of 27 (age range 20-45), from the University of Bristol (students and staff), were compensated for their participation in the experiments. All of them were tested for normal or corrected-to normal vision. Consent forms were signed by each participant and the data were anonymized. Responses from the subjects were first recorded as quality scores in the range 0-100, as explained earlier. Difference scores were then calculated for each tested sequence and each subject $i$ by subtracting the quality score of the distorted sequence $OS_{dis}$ from its corresponding reference $OS_{ref}$. Difference Mean Opinion Scores (DMOS) and the respective statistics (standard error, confidence intervals, etc) were then obtained for each trial by taking the mean of the difference scores among participants $N$. Particularly, DMOS values for a video $v$ were calculated as follows:
\begin{equation} 
\label{eq:DMOS}
DMOS_v = \frac{1}{N} \sum ^N_{i=1} \left(OS_{\text {ref},i} - OS_{\text {dis},i}\right) \, . \end{equation}
In order to reflect the quality taking into account the relative difference from the score of the hidden reference (in this case the original) for each participant $i$, we subtracted the $DMOS_v$ from the maximum quality score (100).

Further to the score collection, we performed post-screening of the subjects. Following the~\cite{r:BT500} protocols, for each resolution group, we performed outlier rejection on all participant scores. No participants were rejected. Moreover, we calculated the subject bias according the Recommendation~\cite{r:P913} and removed it before performing any statistical analysis. As defined in the recommendation, subject bias is the difference between the average of one subject's ratings and the average of all subjects' ratings for each processed video sequence. To remove subject bias, the recommendation proposed to subtract that its value from each one of subject's ratings.

\section{Results and Discussion}
\label{sec:results} 
This section presents the codec comparison results based on objective and subjective quality assessments of BVI-CC, alongside encoder and decoder complexity assessments. For the objective evaluation, two video quality metrics have been employed: the commonly used Peak-Signal-to-Noise-Ratio (PSNR) and Video Multi-method Assessment Fusion (VMAF) \cite{w:VMAF}. The latter is a machine learning-based video quality metric, which predicts subjective quality by combining multiple quality metrics and video features, including the Detail Loss Metric (DLM) (\cite{j:Li2}), Visual Information Fidelity measure (VIF) (\cite{j:Sheikh}), and averaged temporal frame difference (\cite{w:VMAF}). The fusion process employs a \textnu-Support Vector machine (\textnu-SVM) regressor (\cite{j:Cortes}). VMAF has been evaluated on various video quality databases, and shows improved correlation with subjective scores (\cite{w:VMAF,j:Zhang7,j:Li5}). In this work, VMAF has also been employed to determine optimum resolution for each test rate point and sequence, following the procedure described in Section \ref{sec:config}. The difference between test video codecs in terms of coding efficiency was calculated using the \cite{r:Bjontegaard} Delta (BD) measurements benchmarked against HEVC HM.

For the subjective assessment of the BVI-CC dataset, after following the experimental procedure defined in Section~\ref{sec:test}, the raw opinion scores were collected for each trial in confidentiality and anonymized. The rate-quality curves have been plotted for each test sequence in all three resolution groups (see Section~\ref{ssec:SubjectiveComparison}), where the subjective quality is defined as 100-DMOS (see Eq.(~\ref{eq:DMOS})). Before computing the DMOS values and performing a codec comparison, the subject bias was estimated and removed according to the Recommendation~\cite{r:P913}. A significance test was then conducted using one-way Analysis of Variance (ANOVA) between each paired of codecs on all rate points and sequences.

The subjective data collected was also used to evaluate six popular objective video quality metrics (see Section~\ref{ssec:ObjectiveComparison}), including PSNR, Structural Similarity Index (SSIM) (\cite{j:ssim}), multi-scale SSIM (MS-SSIM) (\cite{c:mssim}), VIF (\cite{j:Sheikh}), Visual Signal-to-Noise Ratio (VSNR) (\cite{j:vsnr}), and VMAF (\cite{w:VMAF}). According to the recommendation the bias was only removed from the opinion scores for the subjective comparison of the codecs. For the objective comparison, the raw opinion scores were utilized. Following the procedure in \cite{r:vqeg}, their quality indices and the subjective DMOS were fitted based on a weighted least-squares approach using a logistic fitting function for three different resolution groups. The correlation performance of these quality metrics was assessed using four correlation statistics, the Spearman Rank Order Correlation Coefficient (SROCC), the Linear Correlation Coefficient (LCC), the Outlier Ratio (OR) and the Root Mean Squared Error (RMSE). The definitions of these parameters can be found in \cite{r:vqeg}.

Finally, the computational complexity of the three tested encoders was calculated and normalised to HEVC HM for Resolution Group A and B (see Section~\ref{ssec:ComplexityComparison}). They were executed on the CPU nodes of a shared cluster, Blue Crystal Phase3, of the \cite{w:BC3}. Each node has 16$\times$2.6 GHz SandyBridge cores and 64GB RAM.

\begin{table*}[ht]
 \caption{Codec comparison results based on PSNR and VMAF quality metrics. Here Bj{\o}ntegaard Delta \cite{r:Bjontegaard} measurements (BD-rate) were employed, and HEVC HM was used as benchmark.}
\centering 
\footnotesize
\vspace{.5cm}
\begin{tabular}{l || c c| c c || c c  |c c  ||c | c}
\toprule
	Resolution Group & \multicolumn{4}{c||}{A (UHD)} & \multicolumn{4}{c||}{B (HD)} & \multicolumn{2}{c}{C (HD-DO)} \\ 
	\cmidrule{1-11}	
Codec	& \multicolumn{2}{c|}{PSNR} & \multicolumn{2}{c||}{VMAF} &\multicolumn{2}{c|}{PSNR} & \multicolumn{2}{c||}{VMAF} & \multicolumn{1}{c|}{PSNR}  & \multicolumn{1}{c}{VMAF}\\ 
		\cmidrule{1-11}	
Sequence\textbackslash BD-rate & AV1 & VTM  & AV1 & VTM   & AV1 & VTM  & AV1 & VTM &AV1 & AV1\\ 	
	\midrule
		\midrule 
	AirAcrobatic 		& -12.1\% 	&	-25.5\%				&-12.0\%			&-28.6\%				& -2.6\%   	&-21.7\%			& 4.2\%   	&-20.0\%		& 13.3\%    &     -0.1\%    \\
		\midrule 
	CatRobot				& -6.2\%  	& -38.0\%  	    &-12.8\%    	&-39.6\%   	  	& -4.0\%   	&-37.7\%		  & -10.4\%   	&-41.2\%		& -2.1\%    &    -11.3\%   \\
		\midrule 
	Myanmar 				& 4.3\% 		& -17.2\% 	 	  &1.3\%				&-21.3\%   	  	& 6.5\%   	&-15.5\%		  & 3.5\%   	&-18.6\%		& 8.4\%   &     5.1\%  \\
		\midrule 
	CalmingWater 		& -15.5\% 	& -21.5\% 	  	&-9.6\% 			&-18.9\%  	  	&-15.7\%   	&-22.6\%		  & -10.2\%   	&-19.6\%		& -13.0\%   &     -10.4\%  \\
		\midrule 
	ToddlerFountain & -6.6\%		& -18.7\%		  	&-2.0\% 			&-17.4\%   	 		&-8.1\%   	&-18.2\%		  & -3.7\%   	&-16.4\%		& -7.8\%   &    -7.3\% \\
		\midrule 
	LampLeaves 			& -6.8\% 		& -26.2\% 	  	&-6.2\% 			&-26.1\%   	 		&-2.8\%   	&-23.7\%		  & -0.4\%   	&-24.8\%		& 6.7\%  &     -1.2\%\\
		\midrule 
	DaylightRoad 		& -3.8\%  	& -38.0\%  	    &-12.4\% 			&-40.3\%   	 	  & -0.9\%   	&-37.6\%		  & -10.4\%   	&-42.4\%		& 0.9\%    &    -9.8\%   \\
		\midrule 
	RedRock 				& -3.5\%  	& -32.5\%  	    &-9.0\% 			&-37.9\%   	 	  & 0.8\%   	&-31.5\%   	  & -6.4\%   	&-37.7\%		& 16.1\%    &    -8.0\%   \\
		\midrule 
	RollerCoaster 	& -15.3\%  	& -39.9\%  	    &-14.5\% 			&-41.7\%     	  & -7.9\%   	&-38.9\%      & -11.5\%   	&-39.8\%		& -6.6\%    &    -13.5\%   \\
			\midrule 
					\midrule 
	Average 						& -7.3\%		& -28.5\%			&-8.6\%				&-30.2\%					& -3.8\%		&-27.5\%			& -5.0\%   	&-28.9\%		& 1.8\% & -6.3\% \\
\bottomrule
\end{tabular} 
\label{tab:results}
\end{table*} 

\subsection{Results based on Objective Quality Assessment}

Table \ref{tab:results} summarises the Bj{\o}ntegaard Delta measurements (BD-rate) of AOM AV1 (for three resolution groups) and VVC VTM (for Resolution Group A and B only) compared with HEVC HM, based on both PSNR and VMAF. For the tested codec versions and configurations, it can be observed that AV1 achieves an average bit rate saving of 7.3\% against HEVC HM for the UHD test content assessed by PSNR, and this figure reduces (3.8\%) at HD resolution. When VMAF is employed for quality assessment, the coding gains of AV1 over HM are slightly higher, averaging 8.6\% and 5.0\% for UHD and HD respectively. Comparing to AV1, VTM provides significant bit rate savings for both HD and UHD test content, with average BD-rate values between -27\% and -30\% for PSNR and VMAF. For resolution group C, where VMAF-based DO was applied for HM and AV1, the coding gain achieved by AV1 is 6.3\% (over HM) assessed by VMAF, while there is a BD-rate  (1.8\%) loss when PSNR is employed. In overall conclusion, the performance of AV1 makes a small improvement over HM on the test content, and both AV1 and HM perform (significantly) worse than VTM.

\begin{figure}[htbp]
\begin{minipage}[b]{0.485\linewidth}
\centering
\centerline{\includegraphics[width=.75\linewidth]{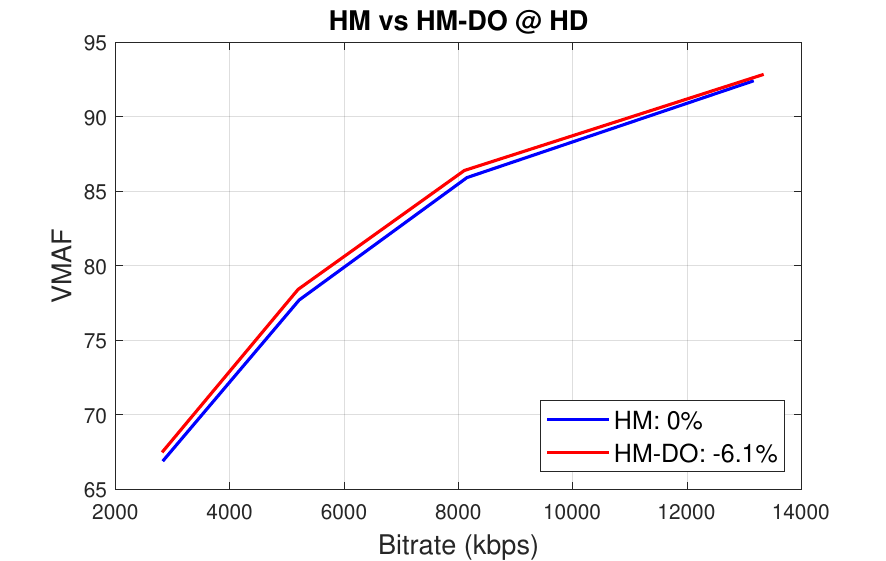}}
\centerline{(a) HM}
\end{minipage}
\begin{minipage}[b]{0.485\linewidth}
\centering
\centerline{\includegraphics[width=.75\linewidth]{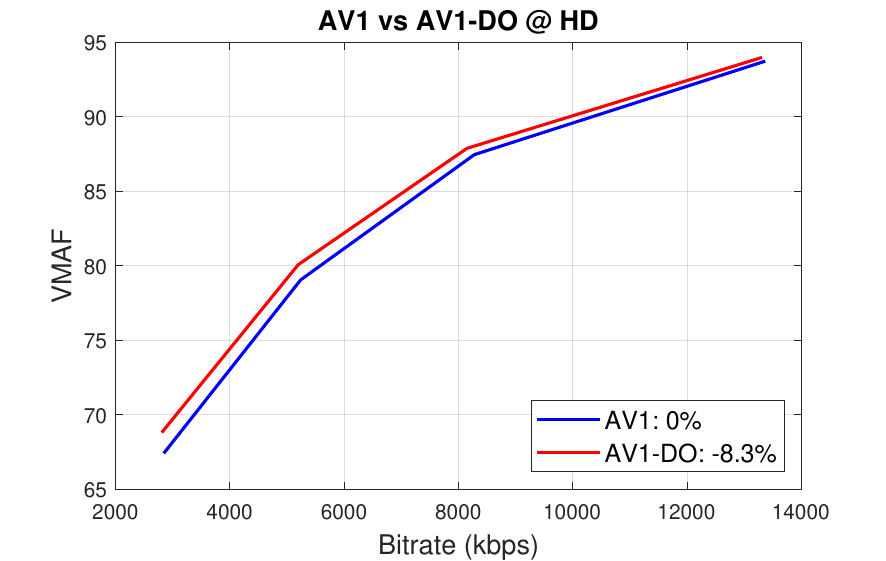}}
\centerline{(b) AV1}
\end{minipage}
\caption{The average rate-VMAF curves of the nine test sequences for HM and AV1 with and without applying DO.}
\label{fig:DO}
\end{figure}

In order to further compare performance in the context of dynamic optimization (DO), the average rate-VMAF curves of the nine test sequences (HD resolution only) for HM and AV1 with and without DO are shown in Fig. \ref{fig:DO}. It can be observed that DO has achieved slightly higher overall coding gains for AV1  (BD-rate is -8.2\%) on the tested content compared to HM (BD-rate is -6.1\%). For both codecs, the savings become lower for higher bit rates (low QP and high quality). It should be noted that the DO approach employed was based on efficient up-sampling using simple spatial filters. \cite{j:Zhang9}, \cite{j:Zhang12} and others have reported significant improvement when advanced up-sampling approaches are applied, such as deep learning based super-resolution.

\subsection{Results based on Subjective Quality Assessment}
\label{ssec:SubjectiveComparison}
\begin{figure*}[htbp]
\scriptsize
\begin{minipage}[b]{0.33\linewidth}
\centering
\centerline{\includegraphics[width=1.0\linewidth]{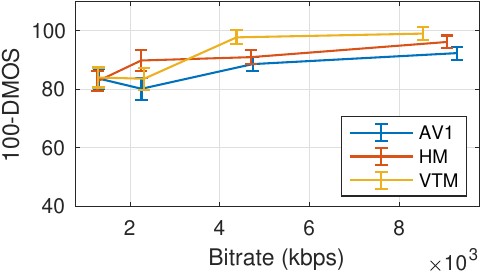}}
\centerline{(a) V1: AirAcrobatic}
\vspace{1em}
\end{minipage}
\begin{minipage}[b]{0.33\linewidth}
\centering
\centerline{\includegraphics[width=1.0\linewidth]{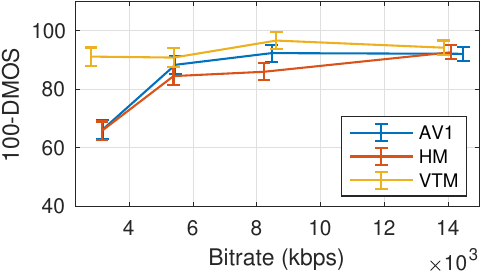}}
\centerline{(b) V2: CatRobot}
\vspace{1em}
\end{minipage}
\begin{minipage}[b]{0.33\linewidth}
\centering
\centerline{\includegraphics[width=1.0\linewidth]{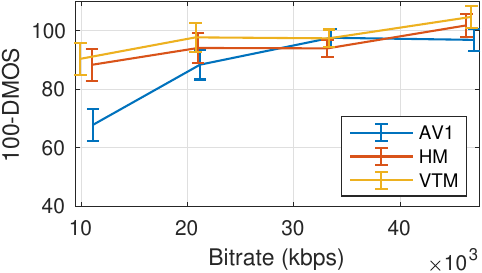}}
\centerline{(c) V3: Myanmar}
\vspace{1em}
\end{minipage}
\begin{minipage}[b]{0.33\linewidth}
\centering
\centerline{\includegraphics[width=1.0\linewidth]{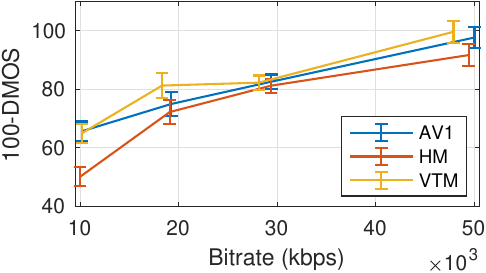}}
\centerline{(d) V4: CalmingWater}
\vspace{1em}
\end{minipage}
\begin{minipage}[b]{0.33\linewidth}
\centering
\centerline{\includegraphics[width=1.0\linewidth]{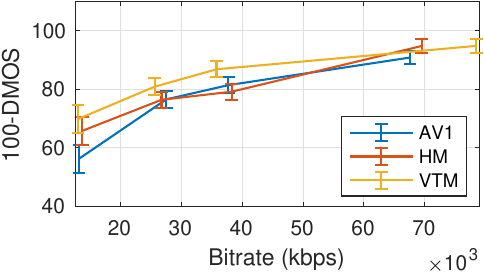}}
\centerline{(e) V5: ToddlerFontain}
\vspace{1em}
\end{minipage}
\begin{minipage}[b]{0.33\linewidth}
\centering
\centerline{\includegraphics[width=1.0\linewidth]{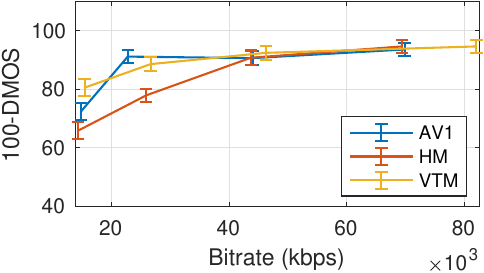}}
\centerline{(f) V6: LampLeaves}
\vspace{1em}
\end{minipage}
\begin{minipage}[b]{0.33\linewidth}
\centering
\centerline{\includegraphics[width=1.0\linewidth]{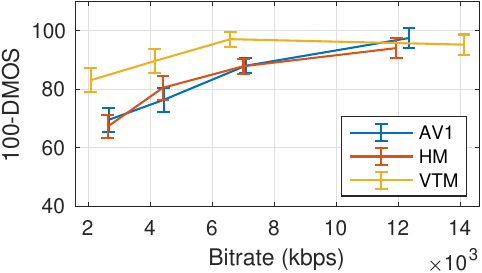}}
\centerline{(g) V7: DaylightRoad}
\end{minipage}
\begin{minipage}[b]{0.33\linewidth}
\centering
\centerline{\includegraphics[width=1.0\linewidth]{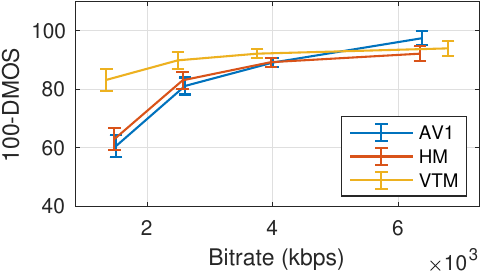}}
\centerline{(h) V8: RedRock}
\end{minipage}
\begin{minipage}[b]{0.33\linewidth}
\centering
\centerline{\includegraphics[width=1.0\linewidth]{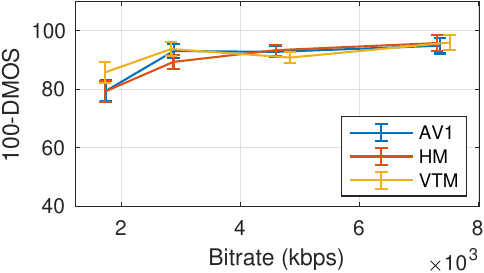}}
\centerline{(i) V9: RollerCoaster}
\end{minipage}
\caption{The DMOS-Rate curves for the Resolution Group A and all three codecs along with the standard error bars after compensating for subject bias, as described in Recommendation \cite{r:P913}.}
\label{fig:DMOSvsRate_4K}
\end{figure*}

\begin{figure*}[htbp]
\scriptsize
\begin{minipage}[b]{0.33\linewidth}
\centering
\centerline{\includegraphics[width=1.0\linewidth]{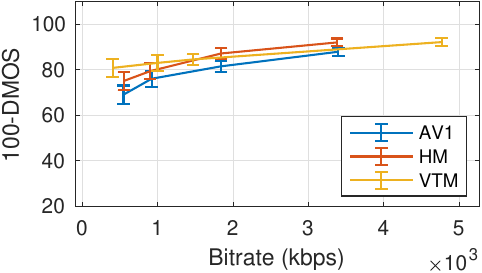}}
\centerline{(a) V1: AirAcrobatic}
\vspace{1em}
\end{minipage}
\begin{minipage}[b]{0.33\linewidth}
\centering
\centerline{\includegraphics[width=1.0\linewidth]{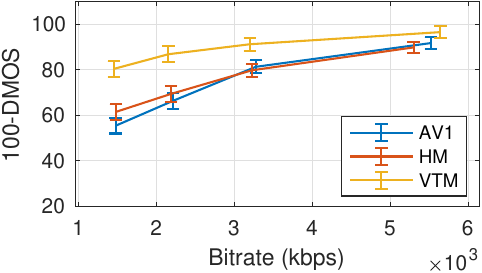}}
\centerline{(b) V2: CatRobot}
\vspace{1em}
\end{minipage}
\begin{minipage}[b]{0.33\linewidth}
\centering
\centerline{\includegraphics[width=1.0\linewidth]{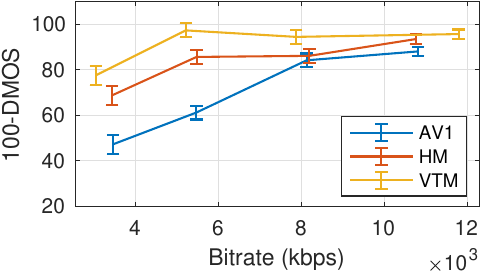}}
\centerline{(c) V3: Myanmar}
\vspace{1em}
\end{minipage}
\begin{minipage}[b]{0.33\linewidth}
\centering
\centerline{\includegraphics[width=1.0\linewidth]{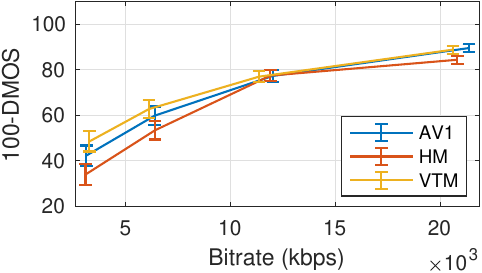}}
\centerline{(d) V4: CalmingWater}
\vspace{1em}
\end{minipage}
\begin{minipage}[b]{0.33\linewidth}
\centering
\centerline{\includegraphics[width=1.0\linewidth]{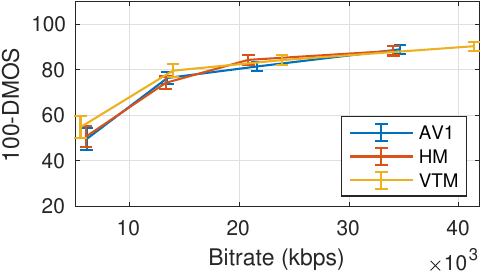}}
\centerline{(e) V5: ToddlerFontain}
\vspace{1em}
\end{minipage}
\begin{minipage}[b]{0.33\linewidth}
\centering
\centerline{\includegraphics[width=1.0\linewidth]{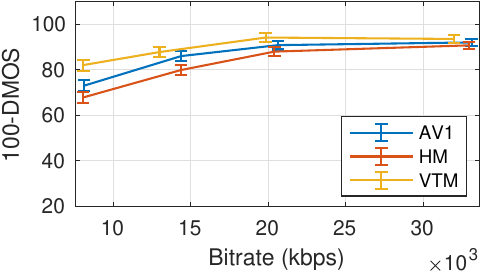}}
\centerline{(f) V6: LampLeaves}
\vspace{1em}
\end{minipage}
\begin{minipage}[b]{0.33\linewidth}
\centering
\centerline{\includegraphics[width=1.0\linewidth]{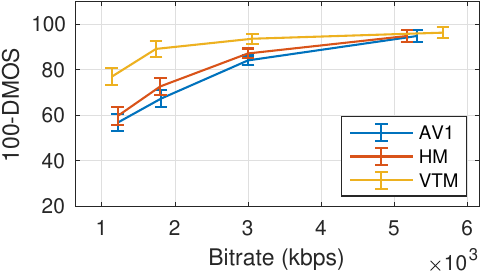}}
\centerline{(g) V7: DaylightRoad}
\end{minipage}
\begin{minipage}[b]{0.33\linewidth}
\centering
\centerline{\includegraphics[width=1.0\linewidth]{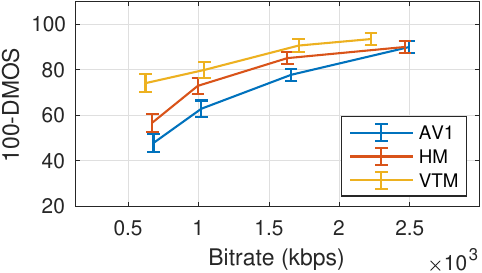}}
\centerline{(h) V8: RedRock}
\end{minipage}
\begin{minipage}[b]{0.33\linewidth}
\centering
\centerline{\includegraphics[width=1.0\linewidth]{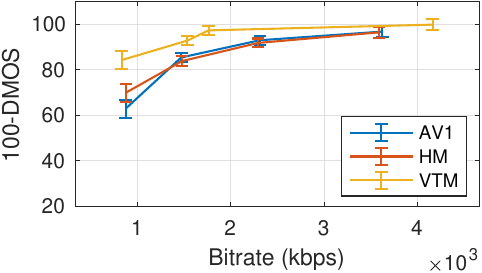}}
\centerline{(i) V9: RollerCoaster}
\end{minipage}
\caption{The DMOS-Rate curves for resolution group B and all three codecs along with the standard error bars after compensating for subject bias, as described in Recommendation \cite{r:P913}.}
\label{fig:DMOSvsRate_HD}
\end{figure*}

As outlined in the introduction of Section~\ref{sec:results}, following the ITU-BT.500 protocols, for each resolution group, we performed outlier rejection on all participant scores. No participants were rejected. Next, following the ITU-T P.913 recommendation~\cite{r:P913}, we estimated the subject bias for the different trials and compensated the raw scores based on that. Figures~\ref{fig:DMOSvsRate_4K}-\ref{fig:DMOSvsRate_HD} illustrate the 100-DMOS against the achieved bit rate for all tested codecs and sequences. Then, we performed one-way ANOVA analysis between pairs of the tested codecs to assess the significance of the differences. Table~\ref{tab:ANOVAscoreUHD1} summarises this comparison.

\par{\textbf{Results on Resolution Group A:}}
 A first impression from Fig.~\ref{fig:DMOSvsRate_4K} is that in most cases the VTM curve is on top of the other curves but also that the confidence intervals are overlapping in most cases. This is confirmed in Table~\ref{tab:ANOVAscoreUHD1}, where the significance tests indicate only four cases which exhibit significant difference ($p < .05$) between HM and AV1. In two of these, AV1 is significantly better than HM: in the cases of CalmingWater sequence at R1 and LampLeaves at R2. Both of these sequences consist of dynamic textures that are very challenging to compress. The opposite happens at R2 for AirAcrobatics and at R1 for Myanmar sequence; two sequences with a static background and slow moving objects. Performing significance tests between VTM and HM, a higher number of cases with significant differences were identified mostly at the lower bit rates. Particularly, for CatRobot at R1, R3; for CalmingWater at R1; for LampLeaves at R1-R2; for DaylightRoad at R1-R3; and for RedRock at R1-R2. 
 A similar number of cases where VTM was significantly better than AV1 were identified: for AirAcrobatics at R3; for CatRobot at R1; for Myanmar at R1; for ToddlerFontain at R1; for DaylightRoad at R1-R3; and for RedRock at R1-R2. It is worth mentioning that from the curves in Fig.~\ref{fig:DMOSvsRate_4K} VTM achieves very good quality ($\geq 80$) even at the lowest bit rates for all sequences except for the ones with dynamic textures (CalmingWater, ToddlerFontain, and LampLeaves). That remains an area for improvement for the future codecs.

\begin{table}[htbp]
\centering
\footnotesize
    \caption{Aggregated significant difference of perceived quality among the tested codecs based on the ANOVA. The first ratio represents the number of sequences where statistically significant difference has been recorded. The ratio in the parentheses show which codec is significantly better (+) or worse (-) in the pairwise comparison of \textit{horizontal/vertical} codec.}
    \vspace{1em}
    \begin{tabular}{c| c c c | c c c| c c }
    \toprule
     &\multicolumn{3}{c|}{\textbf{Resolution Group A}} &\multicolumn{3}{c|}{\textbf{Resolution Group B}} &\multicolumn{2}{c}{\textbf{Resolution Group C}}\\
    \centering 
    \textbf{Codecs} &\textbf{AV1} & \textbf{HM}  & \textbf{VTM}&\textbf{AV1} & \textbf{HM}  & \textbf{VTM}&\textbf{AV1} & \textbf{HM} \\
                  \toprule
     \textbf{AV1} & - & 4/36, (2/-2)& 9/36, (0/-9) & - & 3/36, (0/-3)& 17/36, (0/-17)& - & 5/45, (0/5)\\ 
     \textbf{HM} & 4/36, (2/-2)&- & 10/36, (0/-10)& 3/36, (3/0)&- & 15/36, (0/-15) & 5/45, (-5/0) & - \\ 
     \textbf{VTM} & 9/36, (9/0)& 10/36, (10/0)& -& 17/36, (17/0)& 15/36, (15/0)& -& - & -\\ 
     \bottomrule
\end{tabular}
    \label{tab:ANOVAscoreUHD1}
\end{table}

\par{\textbf{Results on Resolution Group B:}} We performed the same post-screening of the subjects and statistical analysis as in resolution group A.  
Figure~\ref{fig:DMOSvsRate_HD} demonstrates the subjective quality against the bit rate after removing the subject bias and Table~\ref{tab:ANOVAscoreUHD1} summarises the results of the one-way ANOVA comparison of the codecs. Generally, the results for the resolution group B align with those from resolution group A, that in most cases AV1 and HM result in equivalent video quality according to the viewers and that VTM in many cases prevails both codecs. 
Particularly, from Table~\ref{tab:ANOVAscoreUHD1}, it can be observed that HM is significantly better than AV1 in only three cases: for Myanmar at R1-R2 and for RedRock at R2.
VTM is significantly better than HM in 15 cases: for ToddlerFontain at R1-R4; for Myanmar at R1-R2; for CalmingWater at R1; for LampLeaves at R1-R2; for DaylightRoad at R1-R3; at RedRock at R1; and RollerCoaster at R3-R4. Similarly, VTM is significantly better than AV1 in 17 cases (almost have of the test sequences): for AirAcrobatics at R1; for CatRobot at R1-R3; for Myanmar at R1-R4; for LampLeaves at R1; for DaylightRoad at R1-R3; for RedRock at R1-R3; and RollerCoaster at R1-R2. 

\begin{figure}[htbp]
\centering
\begin{minipage}[b]{0.485\linewidth}
\centering
\centerline{\includegraphics[width=\linewidth]{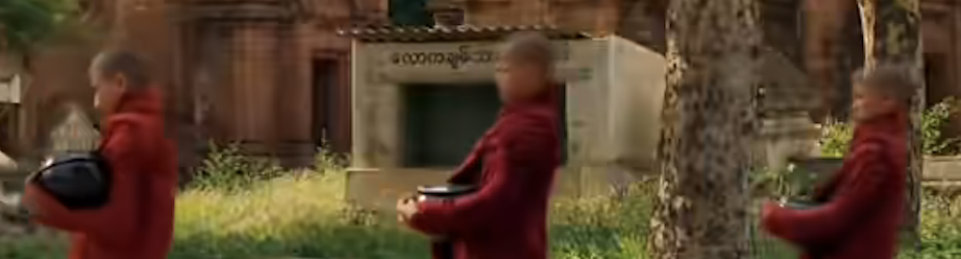}}
\centerline{(a) HM}
\end{minipage}
\begin{minipage}[b]{0.485\linewidth}
\centering
\centerline{\includegraphics[width=\linewidth]{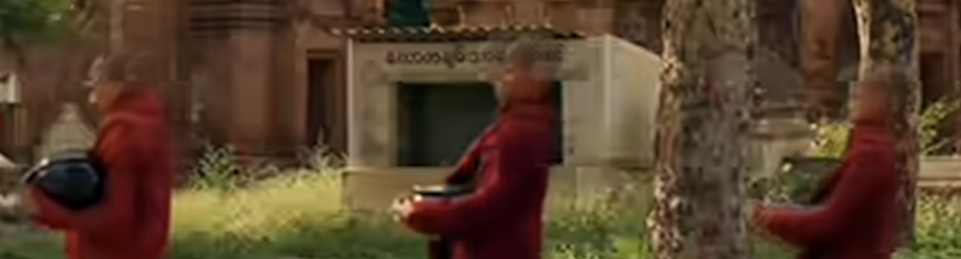}}
\centerline{(b) AV1}
\end{minipage}
\caption{Example of artifacts experienced on the Myanmar sequence that results to subjectively significantly different opinion scores between HM and AV1. The particular patches were captured at R1 from the resolution group B.}
\label{fig:artifactsPatches}
\end{figure}

The reason that significant differences are noticed between AV1 and both HM and VTM in the case of the Myanmar sequence might be associated with the observation that, at lower bit rates, AV1 encoder demonstrates noticeable artifacts on regions of interest, namely in the center of the frame and on the heads of the walking monks in front of a still background with a static camera. An indicative example of these artifacts has been captured in Fig.~\ref{fig:artifactsPatches}. This, however, is probably a rare case as no similar cases have been reported so far in recent literature.

\par
\begin{figure*}[htbp]
\scriptsize
\begin{minipage}[b]{0.33\linewidth}
\centering
\centerline{\includegraphics[width=1.0\linewidth]{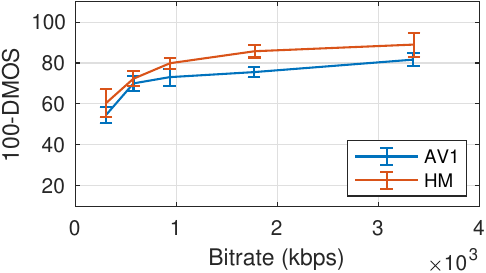}}
\centerline{(a) V1: AirAcrobatic}
\vspace{1em}
\end{minipage}
\begin{minipage}[b]{0.33\linewidth}
\centering
\centerline{\includegraphics[width=1.0\linewidth]{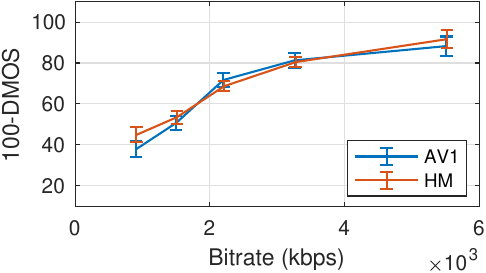}}
\centerline{(b) V2: CatRobot}
\vspace{1em}
\end{minipage}
\begin{minipage}[b]{0.33\linewidth}
\centering
\centerline{\includegraphics[width=1.0\linewidth]{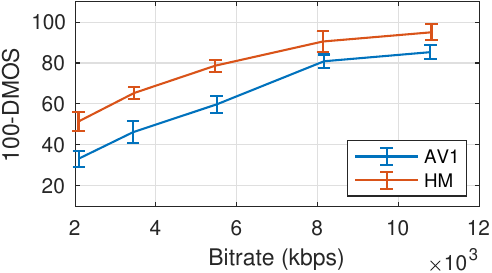}}
\centerline{(c) V3: Myanmar}
\vspace{1em}
\end{minipage}
\begin{minipage}[b]{0.33\linewidth}
\centering
\centerline{\includegraphics[width=1.0\linewidth]{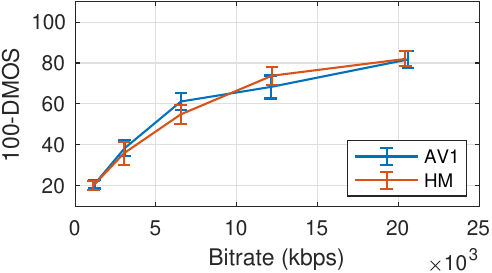}}
\centerline{(d) V4: CalmingWater}
\vspace{1em}
\end{minipage}
\begin{minipage}[b]{0.33\linewidth}
\centering
\centerline{\includegraphics[width=1.0\linewidth]{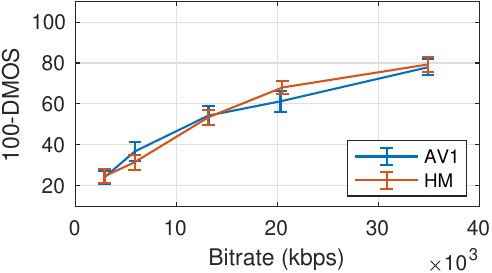}}
\centerline{(e) V5: ToddlerFontain}
\vspace{1em}
\end{minipage}
\begin{minipage}[b]{0.33\linewidth}
\centering
\centerline{\includegraphics[width=1.0\linewidth]{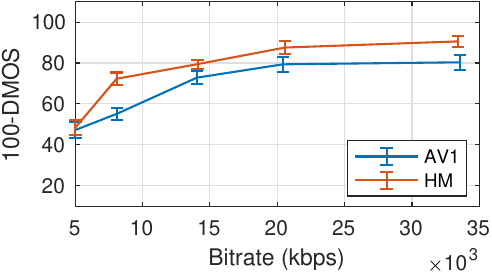}}
\centerline{(f) V6: LampLeaves}
\vspace{1em}
\end{minipage}
\begin{minipage}[b]{0.33\linewidth}
\centering
\centerline{\includegraphics[width=1.0\linewidth]{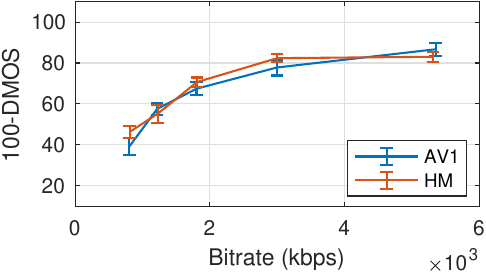}}
\centerline{(g) V7: DaylightRoad}
\end{minipage}
\begin{minipage}[b]{0.33\linewidth}
\centering
\centerline{\includegraphics[width=1.0\linewidth]{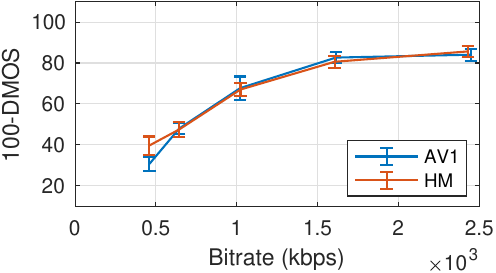}}
\centerline{(h) V8: RedRock}
\end{minipage}
\begin{minipage}[b]{0.33\linewidth}
\centering
\centerline{\includegraphics[width=1.0\linewidth]{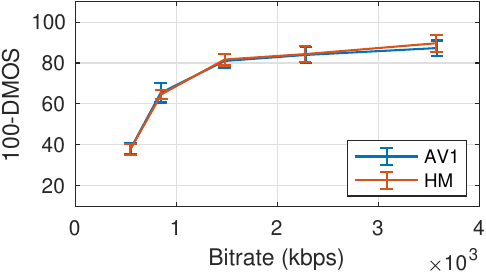}}
\centerline{(i) V9: RollerCoaster}
\end{minipage}
\caption{The DMOS-Rate curves for resolution group C using AV1 and HM codecs along with the standard error bars after compensating for subject bias, as described in Recommendation \cite{r:P913}.}
\label{fig:DMOSvsRate_HDDO}
\end{figure*}

{\textbf{Results on Resolution Group C:}} The removal of the subject bias leads to slightly different results than the ones presented for this case in our previous work (see~\cite{c:Zhang24}), as illustrated in Fig.~\ref{fig:DMOSvsRate_HDDO}. After performing the significance test using one-way ANOVA between paired AV1 and HM sequences, the $p$-values of five rate points were indicated as significantly different. 
In all cases, HM is significantly better than AV1: at R1-R4 for Myanmar; and at R2 and R5 for LampLeaves. Myanmar encoded sequences suffer as mentioned earlier from unique artifacts and LampLeaves is a challenging dynamic texture. Although the findings from this resolution group are generally aligned with the observation from the other two resolution groups, for the LampLeaves sequence we notice a degraded performance of AV1 compared to resolution group B. This is attributed to the selected set of resolutions by the DO algorithm for AV1, which comprises lower resolution sequences than those selected for HM: at R2 540p instead of 720p and at R5 720p instead of 1080p (see Table~\ref{tab:DORes}).

\subsection{Objective Quality Metric Performance Comparison}
\label{ssec:ObjectiveComparison}
The correlation performance of six tested objective quality metrics for three resolution groups (in terms of SROCC values) is summarised in Table \ref{tab:obj}. It can be observed that VMAF outperforms the other metrics on all three test databases with the highest SROCC and LCC values, and lowest OR and RMSE. PSNR results in much lower performance, especially for the UHD resolution group. It is also noted that, for all test quality metrics, the SROCC values for three resolution groups are all below 0.9, which indicates that further enhancement is still needed to achieve more accurate prediction.

\begin{table*}[htbp]
 \caption{The correlation statistics of six popular quality metrics when evaluated on three subject datasets (UHD, HD and HD-DO).}
\centering 
\footnotesize
\vspace{.5cm}
\begin{tabular}{c || cccc || cccc || cccc }
\toprule
Database & \multicolumn{4}{c||}{UHD (108)} & \multicolumn{4}{c||}{HD (108)} & \multicolumn{4}{c}{HD-DO (90)} \\Metric & SROCC & LCC &OR &RMSE & SROCC & LCC &OR &RMSE & SROCC & LCC &OR &RMSE \\
	\midrule
\midrule PSNR & 0.5517&0.6278&0.3056&8.7540 & 0.6097&0.6268&0.5556&12.5870 & 0.7462&0.7439&0.4222&13.3191\\ 
\midrule SSIM & 0.5911&0.5853&0.3148&9.2195 & 0.7194&0.6757&0.4907&11.5968 & 0.8026&0.7836&0.3778&12.2184\\ 
\midrule MSSSIM & 0.7426&0.7436&0.2130&7.4102 & 0.7534&0.7241&0.4537&10.7594 & 0.8321&0.8228&0.3556&11.1398\\ 
\midrule VIF & 0.7464&0.7749&0.1852&6.9273 & 0.7459&0.7592&0.3796&10.0815 & 0.8232&0.8321&0.3778&10.8851\\ 
\midrule VSNR & 0.5961&0.6580&0.2500&8.4062 & 0.5763&0.6587&0.3889&12.0502 & 0.6581&0.7039&0.4778&14.0736\\
\midrule ADM & 0.7532 &0.7639 &0.1759 &7.1573 & 0.6858&0.7290&0.4352&10.8612 & 0.7928&0.8143&0.3556&11.4781\\
\midrule STVMAF &0.7386&0.7471&0.2130&7.3607 & 0.7727&0.7743&0.4167&9.8542 & 0.3147&0.4884&0.5556&17.5840\\
\midrule VMAF & \textbf{0.8463}&\textbf{0.8375}&\textbf{0.1574}&\textbf{5.9972} & \textbf{0.8723}&\textbf{0.8476}&\textbf{0.2870}&\textbf{7.9969} & \textbf{0.8783}&\textbf{0.8840}&\textbf{0.2556}&\textbf{9.1395}\\ 
\bottomrule
\end{tabular} 
\label{tab:obj}
\end{table*}

\subsection{Computational Complexity Analysis}
\label{ssec:ComplexityComparison}
\begin{figure*}[htbp]
\scriptsize
\begin{minipage}[b]{0.245\linewidth}
\centering
\centerline{\includegraphics[width=1.02\linewidth]{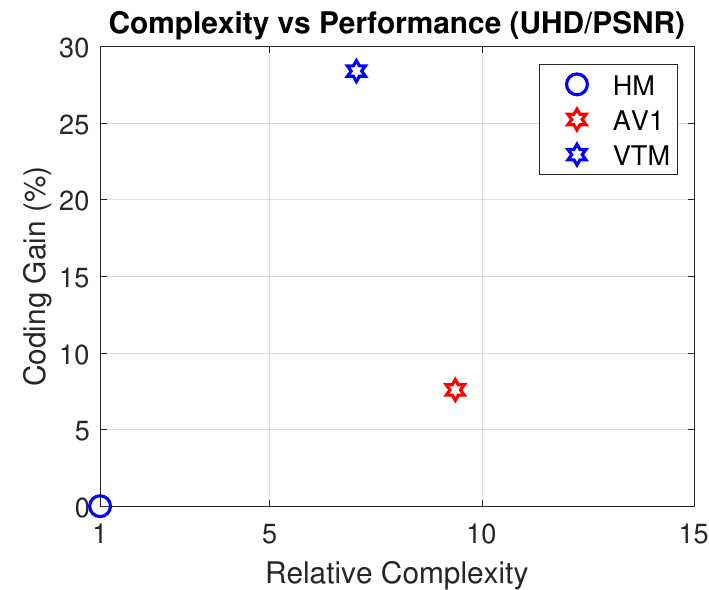}}
\centerline{(a) UHD/PSNR}
\end{minipage}
\begin{minipage}[b]{0.245\linewidth}
\centering
\centerline{\includegraphics[width=1.02\linewidth]{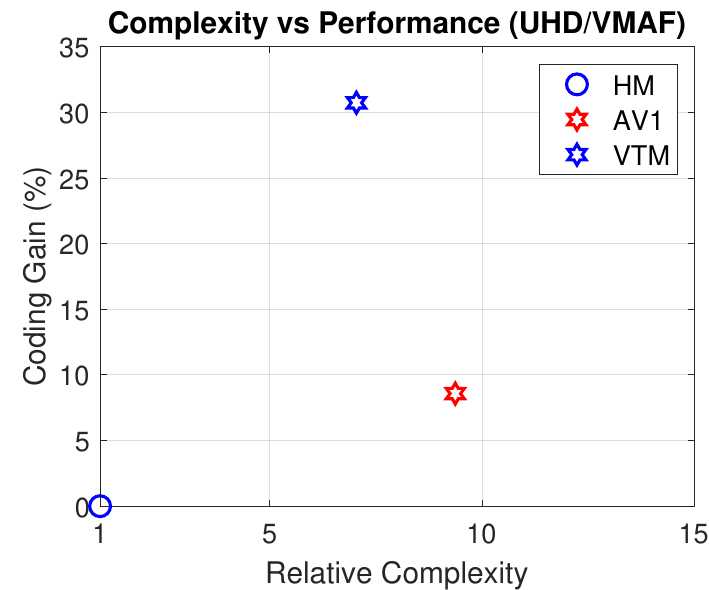}}
\centerline{(b) UHD/VMAF}
\end{minipage}
\begin{minipage}[b]{0.245\linewidth}
\centering
\centerline{\includegraphics[width=1.02\linewidth]{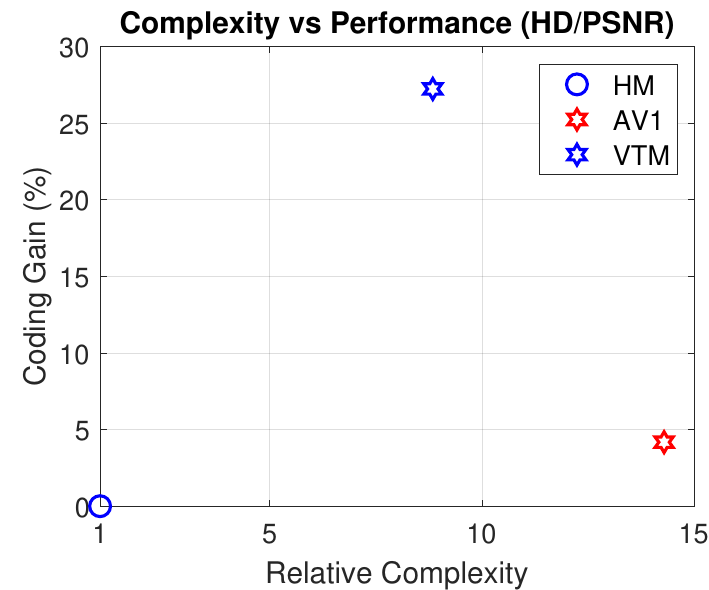}}
\centerline{(c) HD/PSNR}
\end{minipage}
\begin{minipage}[b]{0.245\linewidth}
\centering
\centerline{\includegraphics[width=1.02\linewidth]{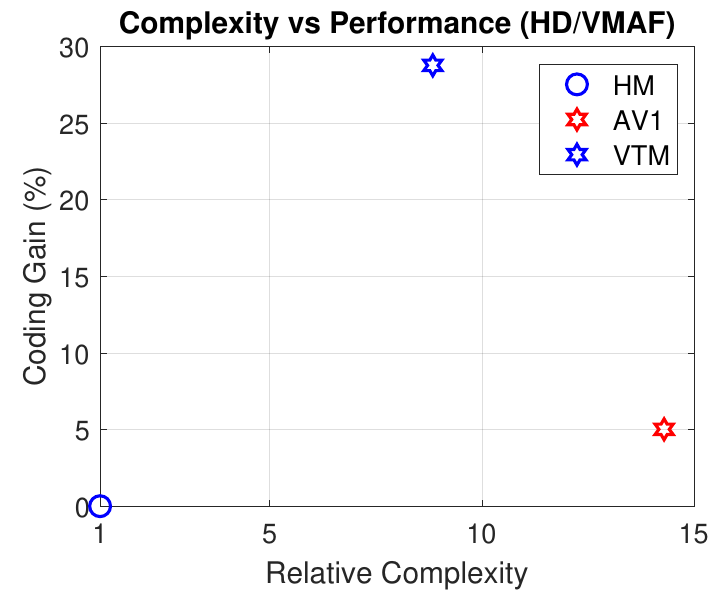}}
\centerline{(d) HD/VMAF}
\end{minipage}
\caption{The relationship between the relative codec complexity (benchmarked on HM) and encoding performance (in terms of average coding gains) for different resolution groups and quality metrics.}
\label{fig:complexity}
\end{figure*}

The average complexity figures for encoding UHD and HD content are summarised in Table \ref{tab:complexity}, where the HM encoder has been used for benchmarking. The average complexity is computed as the average ratio of the execution time of the tested codec for all rate points over the benchmark. As can be seen, for the tested codec versions, AV1 has a higher complexity compared to VTM\footnote{It is noted that the complexity for both AV1 and VTM in more recent versions have been significantly reduced.}. Interestingly these figures are higher for the HD than the UHD resolution. The relationship between the relative complexity and encoding performance (in terms of average coding gains for PSNR and VMAF) is also shown in Fig. \ref{fig:complexity}.

\begin{table}[ht]
 \caption{Computational complexity comparison.}
\centering 
\footnotesize
\vspace{.5cm}
\begin{tabular}{l | c c c}
\toprule
	Resolution Group/Codecs & HM & AV1 & VTM \\ 
	\midrule
	\midrule
	Resolution Group A (UHD)& 1 & 9.37$\times$ & 7.04$\times$\\
	\midrule
	Resolution Group B (HD) &1 & 14.29$\times$ & 8.84$\times$\\
\bottomrule
\end{tabular} 
\label{tab:complexity}
\end{table}

\section{Conclusions}
\label{sec:conclusions} 
This paper presents a video database of nine representative UHD source sequences and 306 compressed versions of these along with their associated objective and subjective quality assessment results. The testing configurations include spatial resampling (from 540p to 1080p) and encoding by three major contemporary video codecs, HEVC HM, AV1, and VVC VTM at pre-defined target bit rates. For one of the three test cases, the convex hull rate-distortion optimisation has been employed to compare HEVC HM and AV1 across different resolutions (from 540p to 1080p) and across a wider bit rate range. 
All the original and compressed video sequences and their corresponding quality scores are available online for public testing (see~\cite{BVI-CC}) with the aim to facilitate research on video compression and video quality across video codecs. To the best of our knowledge, this is the first public dataset that contains encodings from VVC, AV1, and HEVC.

As research on video technologies evolves to data-greedy algorithms, in the near future, we intend to extend this dataset by incorporating more sequences, at higher spatio-temporal resolutions and bitdepths, and by running a large-scale subjective evaluation through crowdsourcing. Furthermore, we intend to extend the set of codecs by including optimized versions of existing standards, such as SVT-AV1 from~\cite{SVT-AV1}, VVenC/VVdeC from~\cite{VVenC}, etc.

\section*{Conflict of Interest Statement}
The authors declare that the research was conducted in the absence of any commercial or financial relationships that could be construed as a potential conflict of interest.

\section*{Author Contributions}
\begin{itemize}
    \item Angeliki Katsenou: Conceptualization, Investigation, Methodology, Subjective Tests, Data Curation, Software, Visualization, Validation, Writing - Reviewing and Editing
    \item Fan Zhang: Conceptualization, Investigation, Methodology, Data Curation, Software, Visualization, Validation, Writing - Reviewing and Editing
    \item Mariana Afonso: Investigation, Methodology, Conceptualization, Software
    \item Goce Dimitrov: Subjective Tests, Data Curation, Software, Validation
    \item David Bull: Supervision, Funding acquisition, Conceptualization, Writing - Reviewing and Editing
\end{itemize}

\section*{Funding}
The authors acknowledge funding from the UK Engineering and Physical Sciences Research Council (EPSRC, project No. EP/M000885/1), the Leverhulme Early Career Fellowship, and the support from NVIDIA Corporation for the donation of GPUs.


\section*{Data Availability Statement}
The data generated during this work and the collection of the test sequences for this study can be found in the BVI-CC (\cite{BVI-CC}) repository.

\bibliographystyle{frontiersinSCNS_ENG_HUMS}
\bibliography{MyRefs.bib}

\end{document}